\documentclass[12pt]{article}
\pdfoutput=1

\usepackage{amsmath}
\usepackage{amsfonts}
\usepackage{amssymb}

\usepackage[
      colorlinks=true,
      linkcolor=blue,
      urlcolor=blue,
      filecolor=black,
      citecolor=red,
      pdfstartview=FitV,
      pdftitle={},
        pdfauthor={Simon A. Gentle, Michael Gutperle, Chrysostomos Marasinou},
        pdfsubject={},
        pdfkeywords={},
        pdfpagemode={},
        bookmarksopen=true
      ]{hyperref}

\usepackage{color}
\usepackage{graphicx}
\usepackage{sectsty}
\sectionfont{\large}

\renewcommand{\thefootnote}{\fnsymbol{footnote}}
\renewcommand{\thanks}[1]{\footnote{#1}}
\newcommand{\starttext}{
\setcounter{footnote}{0}
\renewcommand{\thefootnote}{\arabic{footnote}}}

\newcommand{\no}{\nonumber}

\numberwithin{equation}{section}

\def\cI{{\cal I}}

\def\p{\partial}

\DeclareMathOperator{\tr}{tr}
\usepackage{ dsfont } 

\DeclareMathOperator{\sgn}{sgn}

\renewcommand{\Im}{\operatorname{Im}}
\renewcommand{\Re}{\operatorname{Re}}

\def\no{\nonumber}

\long\def\symbolfootnote[#1]#2{\begingroup%
\def\thefootnote{\fnsymbol{footnote}}\footnote[#1]{#2}\endgroup}

\begin{document}
\setlength{\baselineskip}{18pt}

\starttext
\setcounter{footnote}{0}

\begin{center}

{\Large \bf   Entanglement entropy of Wilson surfaces from\\[0.2cm] bubbling geometries in M-theory}

\vskip 0.4in

{\large  Simon A.\ Gentle, Michael Gutperle and  Chrysostomos Marasinou}

\vskip .2in

{ \it Department of Physics and Astronomy }\\
{\it University of California, Los Angeles, CA 90095, USA}\\[0.5cm]
\href{mailto:sgentle@physics.ucla.edu}{\texttt{sgentle}}\texttt{, }\href{mailto:gutperle@physics.ucla.edu}{\texttt{gutperle}}\texttt{, }\href{mailto:cmarasinou@physics.ucla.edu}{\texttt{cmarasinou@physics.ucla.edu}}

\end{center}

\bigskip

\bigskip
 
\begin{abstract}

\setlength{\baselineskip}{18pt}

We consider  solutions of eleven-dimensional supergravity constructed in \cite{D'Hoker:2008wc,D'Hoker:2008qm}  that are half-BPS, locally asymptotic to $AdS_7\times S^4$ and are the holographic dual of  heavy Wilson surfaces in the six-dimensional $(2,0)$ theory. Using these bubbling solutions we calculate the holographic entanglement entropy for a spherical entangling surface in the presence of a planar Wilson surface. In addition, we calculate the holographic stress tensor and, by evaluating the on-shell supergravity action,  the expectation value of the Wilson surface operator.

\end{abstract}

\setcounter{equation}{0}
\setcounter{footnote}{0}

\newpage

\tableofcontents

\newpage
\section{Introduction}
\label{sec1}

The celebrated proposal of Ryu and Takyanagi \cite{Ryu:2006bv,Ryu:2006ef} provides a method to calculate entanglement entropies  for field  theories that have holographic duals. 
Originally the proposal was used to calculate the entanglement entropy for theories in their vacuum state, but was quickly generalized to include more general settings, such as finite temperature and time-dependent states (see e.g.\ \cite{Nishioka:2009un} for a review). For spherical  entangling surfaces it was observed by Casini, Huerta and Myers \cite{Casini:2011kv} that the holographic  entanglement entropy can be mapped to the thermal entropy of a hyperbolic black hole. In the field theory, the corresponding entanglement entropy  is  mapped to the  thermal entropy on a hyperbolic space. 

 Another generalization concerns entanglement entropy in the presence of extended defects,  such as Wilson lines.  
 In the  probe approximation these defects are described by strings or branes in the $AdS$ space and the backreaction  of the stress energy of the branes on the bulk geometry is neglected.
  For example, the holographic description of a Wilson loop in $SU(N)$  ${\cal N}=4$ SYM in the fundamental representation is given by a fundamental string in $AdS_5\times S^5$ \cite{Rey:1998ik,Maldacena:1998im}, whereas higher dimensional representations can be described by D3- (D5-) branes with $AdS_2 \times S^2 (S^4)$ worldvolume in $AdS_5\times S^5$ \cite{Gomis:2006sb,Gomis:2006im}. See \cite{Chang:2013mca,Jensen:2013lxa, Estes:2014hka} for a discussion of entanglement entropy in the presence of probe brane defects.  
  When the dimension of the representation increases and becomes of order $N^2$, the backreaction cannot be neglected and the probe is replaced by a new bubbling geometry with flux. The bubbling holographic solutions corresponding to half-BPS Wilson loops in ${\cal N}=4$ SYM were found in \cite{D'Hoker:2007fq,Lunin:2006xr,Yamaguchi:2006te}.
  
  In  \cite{Lewkowycz:2013laa}, Lewkowycz and Maldacena applied the  Casini, Huerta and Myers  mapping to the calculation of the entanglement 
  entropy in the presence of  Wilson loops in ${\cal N}=4$ SYM theory and ABJM theories \cite{Aharony:2008ug}. They  showed that the entanglement entropy can be calculated from 
  the expectation value of the Wilson loop operator as well as the one point function of the stress tensor in the presence of the Wilson loop. For BPS Wilson 
  loops these quantities can be evaluated using localization and reduced to matrix models \cite{Pestun:2007rz,Kapustin:2009kz}.  In  \cite{Gentle:2014lva} by two of the present authors, the holographic entanglement entropy was calculated using the  bubbling solution dual to half-BPS Wilson loops.  It was shown that the result agrees with  \cite{Lewkowycz:2013laa} when the exact map between matrix 
 model quantities and the  supergravity solution, found in \cite{Okuda:2008px,Gomis:2008qa}, is applied.

The goal of the present paper is to generalize the holographic calculation of the entanglement entropy to  the case  of six-dimensional $(2,0)$ theory with 
half-BPS Wilson surfaces present. The $(2,0)$ theory can be defined  either by a low-energy  limit of   type IIB string theory on an $A_{N-1}$ singularity \cite{Witten:1995zh} or by a decoupling limit  of $N$ coincident M5 branes \cite{Strominger:1995ac}.  While there exists no simple  
Lagrangian description  of this theory due to the presence of tensor fields $B^+$ with self-dual field strength,  the holographic dual  \cite{Maldacena:1997re} is given by M-theory on $AdS_7\times S^4$. In analogy with the Wilson loop one expects that the six-dimensional theory has extended Wilson surface operators \cite{Ganor:1996nf} of the form
\begin{align}
W_\Gamma&\sim\tr \exp\left( \int_\Gamma B^+\right)
\end{align}

In the probe approximation the Wilson surface operators can be described by embedding  M2-branes  \cite{Berenstein:1998ij,Corrado:1999pi}  or M5-branes  \cite{Chen:2007ir,Chen:2008ds,Mori:2014tca} on various submanifolds inside the $AdS_7\times S^4$.   It is an interesting open question whether the expectation value of Wilson surface operators can be calculated by localization in the $(2,0)$ theory.

\begin{figure}[!t]
  \centering
  \includegraphics[width=60mm]{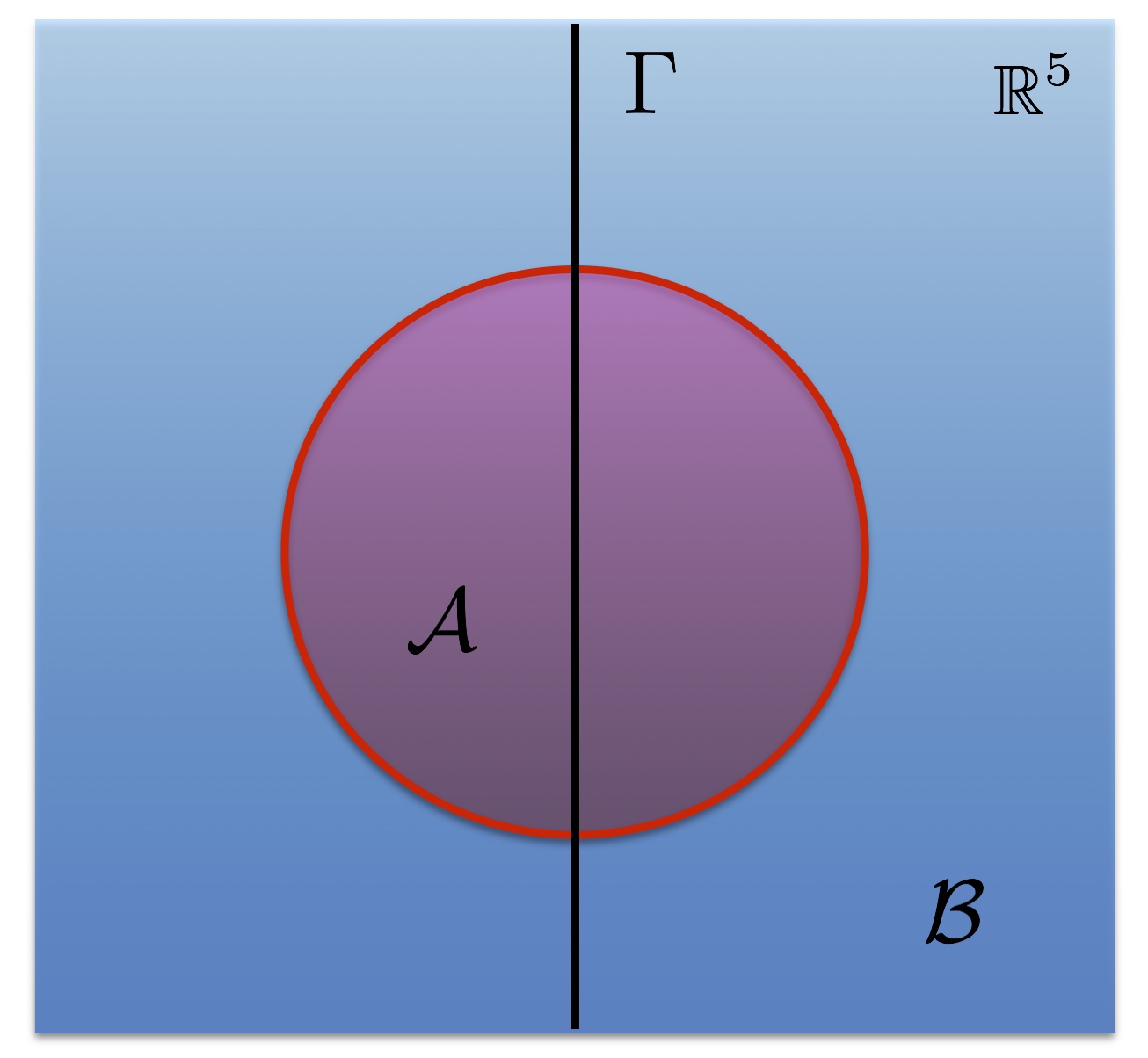}
  \caption{The spherical entangling surface $\partial {\cal A}$ is the boundary of a region ${\cal A}$ on a constant time slice of the $(2,0)$ theory on $\mathbb{R}^{6}$.   The  Wilson surface $\Gamma$ intersects this surface twice.}
  \label{fig:entanglesurf}
\end{figure}

In \cite{D'Hoker:2008wc,D'Hoker:2008qm}, bubbling solutions corresponding to half-BPS Wilson surfaces were found (see \cite{Lunin:2007ab} for earlier work in this direction). In the present paper we will use these  solutions of eleven-dimensional supergravity   to calculate the entanglement entropy as well as other holographic observables.  The bubbling solutions feature an $AdS_3\times S^3\times S^3$ fibration over a two-dimensional Riemann surface $\Sigma$ with boundary.  The solutions are locally asymptotic to $AdS_7\times S^4$ and the six-dimensional  asymptotic metric on the $AdS_7$ boundary is naturally $AdS_3\times S^3$. It is  convenient   to describe the Wilson surface on this space by imposing boundary conditions at the boundary of $AdS_3$ and choosing Poincar\'{e} coordinates for the $AdS_3$ factor describes a planar Wilson surface. However, as  discussed in section~\ref{sec33}, this metric on $AdS_3\times S^3$ can be related to the  more familiar flat metric on  $\mathbb{R}^{6}$ by a conformal transformation. It is easier to visualize the geometry of our setup on $\mathbb{R}^{6}$: the entangling surface at constant time  is a four-sphere of radius $R$ and the Wilson surface  is a line (also filling out the time direction) that intersects the four-sphere at two points, as illustrated in figure~\ref{fig:entanglesurf}.

\subsection{Summary of results}

For the convenience of the reader we will present our main results here and put them into context. The change of the entanglement entropy due to the presence of the Wilson surface is given by
 \begin{align}
 \Delta S_{\cal A}&=  \frac{ 4 N^3}{3} \, \left[  \frac{16+3  m_2^2-8 m_3}{320}
 - \frac{1}{16 } \sum_{i<j} (-1)^{i+j} |\xi_i-\xi_j|^3 -\frac{1}{2}\right] \log\left(\frac{2R }{\eta}\right)\label{eq:EEresult}
\end{align}
We also calculate two other holographic observables.  The stress tensor in the presence of the Wilson surface for the $AdS_3\times S^3$ boundary coordinates  is given by
 \begin{align}
 \Delta \langle T_{i j} \rangle\, dx^{i} dx^{j} &= \frac{N^3}{160 \pi^3} \left(16 + 3m_2^2 - 8m_3\right)   \left( ds^2_{AdS_3} - ds^2_{S^3} \right) 
\end{align}
Finally we calculate the expectation value of the Wilson surface operator by evaluating the regularized on-shell supergravity action:
\begin{align}
\log\, \langle W_{\Gamma} \rangle &= - \frac{N^3}{192 \pi}\, \textrm{Vol}(AdS_3) \left({\cal F}+64\right)
\end{align}
The  $m_{2,3}$ are quantities that depend on the parameters $\xi_i$ of a general bubbling solution. The cut-off $\eta$ is the distance from the Wilson surface, as discussed in more detail in section~\ref{sec33}. Also, ${\cal F}$ is a finite one-dimensional integral that is defined in section~\ref{sec532}.

\subsection{Structure of the paper}
The structure of the paper is as follows. In section~\ref{sec2} we review the bubbling  half-BPS solutions of M-theory originally obtained in  \cite{D'Hoker:2008wc,D'Hoker:2008qm} and work out  the behavior of the solution near the asymptotic boundary. In particular, we  determine the Fefferman-Graham map for an  asymptotic $AdS_3\times S^3$ boundary metric. In section~\ref{sec3} we calculate the entanglement entropy for a spherical  entangling surface following the Ryu-Takayanagi prescription for the bubbling solution. In section~\ref{sec4} we use the methods of Kaluza-Klein holography and holographic renormalization to calculate the one point function of the stress tensor for the bubbling solution. In section~\ref{sec5}  we evaluate the on-shell action of eleven-dimensional supergravity to determine the expectation value of the Wilson surface. We show that the bulk part of the action is given by a total derivative and evaluate the integral as well as the Gibbons-Hawking term. In section~\ref{sec6} we provide a brief discussion of our results as well as possible avenues for future research. In the interest of readability we present  technical matters and details of calculations in several appendices.

\section{Review of bubbling M-theory solutions}
\setcounter{equation}{0}
\label{sec2}

In this section we will review the  construction of half-BPS   M-theory solutions found in \cite{D'Hoker:2008wc} that are locally asymptotic to $AdS_7\times S^4$. These solutions generalize  the construction of Janus solutions \cite{Bak:2003jk, D'Hoker:2007xy} in type IIB   to M-theory. They  correspond to the holographic description of Wilson surface defects in the six-dimensional $(2,0)$ theory, where the Wilson surface is `heavy' and the backreaction on the geometry  is taken into account. 

One demands that these solutions preserve an $OSp(4^*|2)\oplus OSp(4^*|2)$ sub-superalgebra of the $OSp(8^*|4)$ superalgebra of the  $AdS_7\times S^4$ vacuum.  This form of the preserved superalgebra  is uniquely determined by demanding that the solution has sixteen unbroken supersymmetries and preserves $so(2,2|\mathbb{R})$ associated with conformal symmetry on the worldvolume of the Wilson surface, $so(4|\mathbb{R})$ corresponding to rotational symmetry in the space transverse to the Wilson surface and an unbroken   $so(4|\mathbb{R})$ R-symmetry \cite{D'Hoker:2008ix}. Note that  a generalization was recently analyzed  in  \cite{Bachas:2013vza} in which the preserved superalgebra is $D(2|1,\gamma)\oplus D(2|1,\gamma)$, but we will not discuss this case here.

 It follows from these superalgebra considerations that the bubbling BPS solution has an  $so(2,2|\mathbb{R})\oplus so(4|\mathbb{R}) \oplus so(4|\mathbb{R})$  algebra of isometries. Furthermore, the  solution preserves sixteen of the thirty-two supersymmetries.  The BPS equations were solved in \cite{D'Hoker:2008wc} and the global regular solutions were found in \cite{D'Hoker:2008qm}.  The ansatz for the eleven-dimensional metric is given by an $AdS_3 \times S^3\times S^3$ fibration over a Riemann surface $\Sigma$ with boundary:
 \begin{align}
 \label{metans}
ds^2&= f_1^2 \, ds^2_{AdS_3}+ f_2^2\,  ds_{S^3}^2+ f_3^2 \, ds_{\tilde S^3}^2+ 4 \rho^2\,  |dv|^2
\end{align}
where we denote the complex coordinate of the two-dimensional Riemann surface by $v$. In addition, $ds_{S^3}^2$ and $ds_{\tilde S^3}^2$ are the metrics on the unit-radius three-spheres and the metric on the unit-radius Euclidean $AdS_3$ in Poincar\'e half-plane coordinates is given by
\begin{align}
\label{eq:AdS3metric}
ds^2_{AdS_3}& = \frac{dz^2+dt^2+dl^2}{z^2}
\end{align}
The Wilson surface on the boundary $AdS_3\times S^3$ fills the $t,l$ directions and is located at $z=0$.

\begin{figure}
  \centering
  \includegraphics[width=100mm]{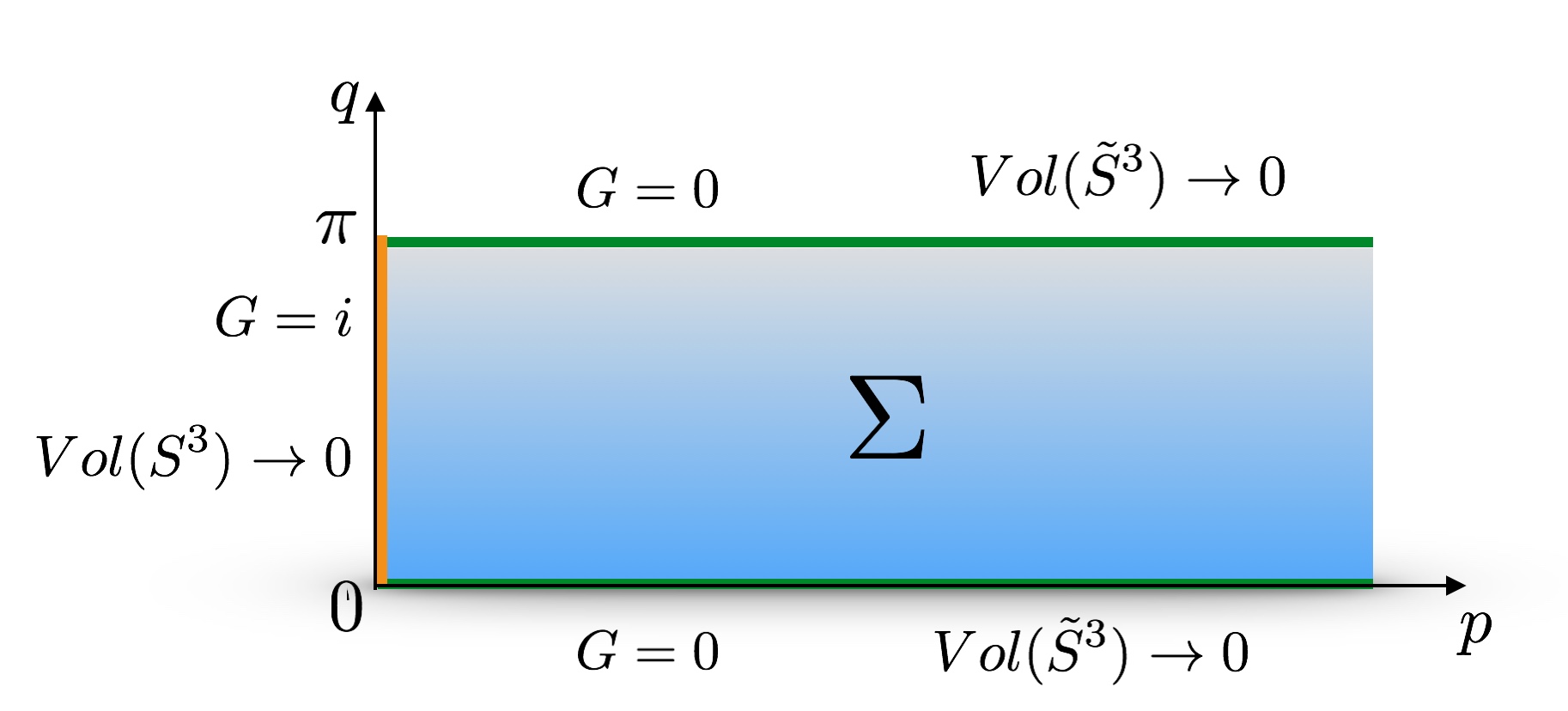}
  \caption{$AdS_7\times S^4$ parameterized on the half strip.}
  \label{fig:vacuumgeometry}
\end{figure}

These solutions are parametrized by a harmonic function $h$ and a complex function $G(v,\bar v)$ that satisfies a first order differential equation:
\begin{align}
\label{diffeqa}
\partial_v G&= {1\over 2} \left(G+\bar G\right) \partial_v \ln h
\end{align}
It is useful to introduce the combinations\footnote{Note that there is a typo in eq (2.5)  of \cite{D'Hoker:2008qm}:  it should read $W^2= -4 |G|^2-(G-\bar G)^2$.}
\begin{align}
W_+ = |G-\bar G| +2 |G|^2, \quad  W_- = |G-\bar G| -2 |G|^2
\end{align}
in terms of which the  metric functions in (\ref{metans}) are given by
\begin{gather}
f_1^6 =  4h^2 (1-|G|^2)\, {W_+ \over W_-^2}, \quad f_2^6=  4h^2 (1-|G|^2)\, {W_- \over W_+^2} \nonumber \\
f_3^6= {h^2 W_+ W_-\over  16 (1-|G|^2)^2}, \quad \rho^6 = {(\partial_v h \partial_{\bar v} h)^3\over 16  h^4}\, (1-|G|^2)\, W_+ W_-  \label{met2}
\end{gather}
It was shown in \cite{D'Hoker:2008qm} that for a  solution to be regular  the functions $h$ and $G$ must satisfy the following conditions on the  Riemann surface $\Sigma$  and its boundary:
\begin{align}\label{ads7sol}
h &=0, \;\; G = 0, +i  , \quad \quad   v\in \partial \Sigma \nonumber \\
h&>0,\;\; |G|^2<1  \quad \quad \quad \;  v \in  \Sigma
\end{align}
First we consider the simplest example: the  $AdS_7\times S^4$ vacuum  solution.  This can be obtained by choosing $\Sigma$ to be the half strip $\Sigma=\{ v =p+i q/2, p>0, q\in[0,\pi]\}$ with
\begin{align}
h&= -i L^3 \left( \cosh(2v)-\cosh(2\bar v)\right), \quad  G= -i  \, {\sinh(v-\bar v) \over \sinh 2\bar v}
\end{align}
which produces
\begin{align}
f_1&= 2L\cosh p, \quad f_2=2L\sinh p , \quad f_3 =L\sin q , \quad \rho =L
\end{align}
Hence the $AdS_7\times S^4$ metric is given by
\begin{align}
\label{adsmeta}
ds^2&= 4 L^2 \left( dp^2+\cosh^2 p\, ds_{AdS_3}^2 +  \sinh^2 p\, ds^2_{S^3} \right)  +  L^2  \left( dq^2 + \sin^2 q\, ds^2_{\tilde S^3}\right)
\end{align}
This geometry is represented in figure~\ref{fig:vacuumgeometry}. 
More general bubbling solutions can be constructed once we realize that the $AdS_7\times S^4$ solution can be mapped from the half strip  to the upper half-plane  via
\begin{align}
w&=\cosh(2v) 
\end{align}
The functions $h$ and $G$ then take the form 
\begin{align}
\label{hgdefine}
h&= -iL^3(w-\bar w),\quad \quad G= {i\over 2} \left( {w+1 \over  \sqrt{(w+1)(\bar w+1)}} - {w-1 \over  \sqrt{(w-1)(\bar w-1)}}   \right)
\end{align}
Note that the boundary of $\Sigma$ is now located at the real line and that on the real line  the function $G=+i$ when $\Re w\in[-1,1]$ and $G=0$ when $\Re w >1$ or $\Re w <-1$.

\begin{figure}
  \centering
  \includegraphics[width=150mm]{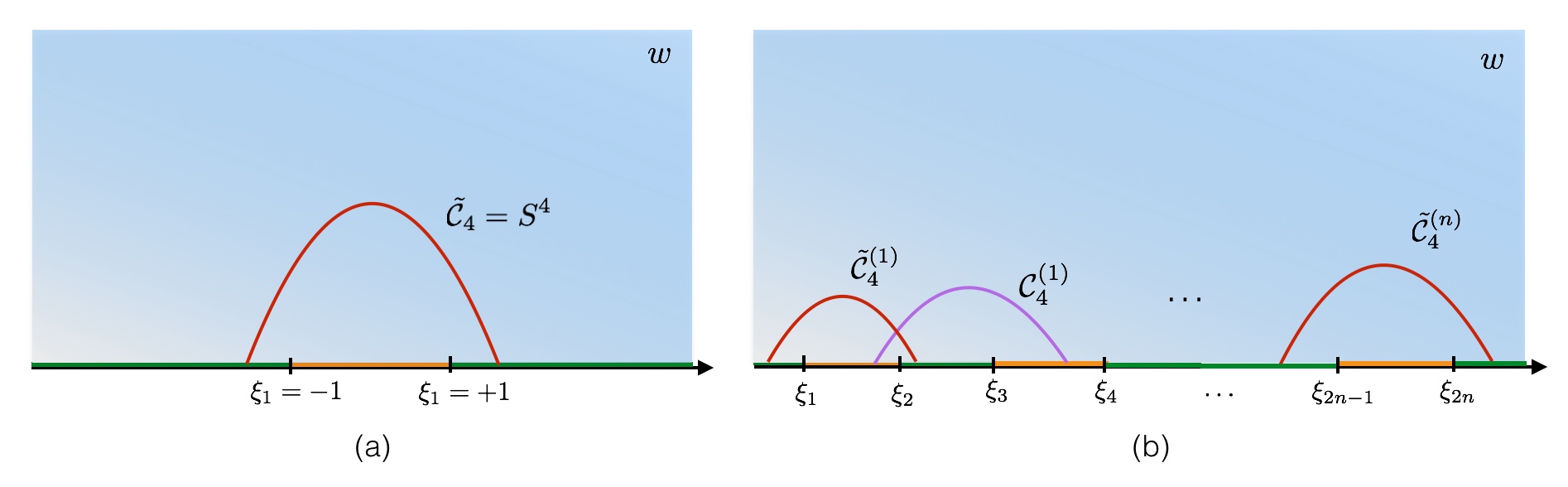}
  \caption{(a) $AdS_7\times S^4$ on the upper half-plane. (b) General bubbling solution with $n$ four-cycles $\tilde {\cal C}_4^{(i)}$, $i=1,\ldots,n$ and $n-1$ four-cycles ${\cal C}_4^{(i)}$, $i=1,\ldots,n-1$.}
  \label{figure3geo}
\end{figure}
A general bubbling solution is constructed by choosing a simple form for $h$ and the following linear superposition for $G$:
\begin{align}
\label{gdefa}
h &= -iL^3(w-\bar w),\quad \quad G= \sum_{i=1}^{2n} (-1)^i g(\xi_i), \quad \quad g(\xi) \equiv  -{i\over 2}  {w-\xi  \over  \sqrt{(w-\xi)(\bar w-\xi)}}
\end{align}
where $w$ is now a general coordinate on the upper half-plane.
The  solution is completely characterized by the choice of $2n$  real numbers $\xi_i$ with $i=1,2,\ldots, 2n$ 
that are ordered 
\begin{align}\label{eq:xiordering}
-\infty=\xi_0<\xi_1< \xi_2<\cdots< \xi_{2n}< \xi_{2n+1}=+\infty
 \end{align}
 We have introduced $\xi_0$ and $\xi_{2n+1}$ to  simplify the expression for the boundary condition that the function $G$ satisfies on the real line:
 \begin{align}
G|_{\Im w=0}= 
\left\{
\begin{array}{cc}
 0 &    \Re w\in[\xi_{2k},\xi_{2k+1}]   \\
 +i  &    \Re w\in[\xi_{2k+1},\xi_{2k+2}]    \\   
\end{array}
\right. , 
\begin{array}{cc}
\quad k=0,1,2,\ldots,n\\
\quad k=0,1,2,\ldots,n-1\\
\end{array}
\end{align}

A general bubbling solution is characterized by the appearance of new nontrivial four-cycles (see figure~\ref{figure3geo}). The $n$ four-cycles $\tilde{\cal C}_4^{(i)}$ are constructed by connecting two boundary points on different intervals where the volume of the three-sphere $\tilde S^3$ shrinks to zero. This generalizes the construction of the four-sphere in the $AdS_7\times S^4$ vacuum solution. In addition, the geometry also has $n-1$ four-cycles ${\cal C}_4^{(i)}$ that are constructed by connecting points on different intervals where the three-sphere $S^3$ shrinks to zero size. In the bubbling solution these cycles carry nontrivial four-form flux and are the remnants of M5-branes wrapping $AdS_3\times S^3$ and $AdS_3\times \tilde S^3$, respectively.

\subsection{Asymptotic behaviour and regularization}
\label{sec:asymreg}

In this section we  study the asymptotic behaviour of a general bubbling solution.  We will see later that  the area integral and the action integral both diverge, so we need to regulate the integrals and map the regulator to the Fefferman-Graham (FG) UV cut-off.

It is convenient to choose the following coordinates on $\Sigma$: $w=   r\, e^{i \theta}$.  The boundary of $AdS_7\times S^4$ is located at $r\to \infty$.  The expressions for $G$ and $\bar G$ given in (\ref{gdefa}) can be expanded at large $r$ in terms of the  generating function of the Legendre polynomials
\begin{align}\label{eq:Legendre}
{1\over \sqrt{1- 2 x t+ t^2}}= \sum_{k=0}^\infty P_k(x)\, t^k
\end{align}
with the result
\begin{align}\label{largerg}
G= {i\over2} \sum_{k=1}^{\infty} \frac{a_k (\theta)\, m_k}{r^k}, \quad \bar G= -{i\over2} \sum_{k=1}^{\infty} \frac{\bar a_k (\theta)\,  m_k}{r^k}
\end{align}
where the dependence on the angular coordinate $\theta$ is given by 
\begin{align}
a_k(\theta)&\equiv P_{k-1}(\cos \theta)-e^{i\theta} P_k(\cos \theta)\nonumber\\
\bar a_k(\theta)&\equiv P_{k-1}(\cos \theta)-e^{-i\theta} P_k(\cos \theta)
\end{align}
The moments $m_k$ are defined via
\begin{align}
\label{momentdefa}
m_k &\equiv \sum_{i=1}^{2n} (-1)^i \xi_i^k
\end{align}
To ensure that a general bubbling solution is asymptotic to $AdS_7\times S^4$  with radii 
\begin{align}
 R_{S^4}= {R_{AdS_7} \over 2}  = L 
\end{align}
we must identify $m_1 \equiv 2$.
This provides a constraint on the $\xi_i$. Also, for the $AdS_7\times S^4$ solution ($n=1$)  we note here  that all even moments vanish and all odd moments equal $2$.

The metric functions take the following forms as power series in large $r$:
\begin{align}
{f_1^2 \over L^2} &= 2 r + {4-m_2 \cos\theta \over 2} + {3(8+m_2^2-2  m_3) +(8 + 3 m_2^2-10  m_3) \cos 2\theta \over 24  \, r}  +O\left({1\over r^2}\right)\nonumber\\
{f_2^2 \over L^2} &= 2 r - {4+m_2 \cos\theta \over 2 } + {3(8+m_2^2-2  m_3) +(8 + 3 m_2^2-10  m_3) \cos 2\theta \over 24 \, r}  +O\left({1\over r^2}\right)\nonumber\\
{f_3^2 \over L^2 \sin^2\theta} &= 1  + {m_2 \cos\theta \over 2\, r}  -{3(m_2^2-8 m_3) +(32  + 3 m_2^2-40  m_3)\cos2\theta  \over  96 \, r^2 } +O\left({1\over r^3}\right)  \nonumber\\
{\rho^2 \over L^2}&= {1\over 4\, r^2} + {m_2\cos\theta\over 8\, r^3}+ {-3(16  +m_2^2-8  m_3) +(16-3 m_2^2+ 40 m_3) \cos2\theta \over 384 \, r^4} + O\left({1\over r^5}\right)\label{fexplgr}
\end{align}
Next we present the mapping of the $(r,\theta)$ coordinates for large values of $r$ to an FG coordinate system $(u, \tilde\theta)$  for a general bubbling solution.  We need this map to define the large $r$ cut-off function as well as to perform the Kaluza-Klein (KK) reduction in the calculation of the expectation value of the stress tensor in section~\ref{sec4}.

It is natural to consider a Wilson surface living on $AdS_3\times S^3$.  This space is related to $\mathbb{R}^{6}$ by a Weyl rescaling.  We can choose to adapt our FG chart to either space; here we choose the former.  The general FG metric that preserves the  $AdS_3 \times S^3\times S^3$ isometry of a bubbling solution is given by
\begin{align}
\label{eq:FG}
ds^2&= L^2\left[ \frac{4}{u^2} \left( du^2 +\alpha_1 ds^2_{AdS_3} +\alpha_2 ds^2_{S^3}   \right) +\alpha_3 d\tilde{\theta}^2 + \alpha_4 ds^2_{\tilde{S}^3}  \right]
\end{align}
 Equating this  metric with the bubbling metric  \eqref{metans} we find
\begin{gather}
f_1^2 = \frac{4L^2 \alpha_1}{u^2}, \quad f_2^2 = \frac{4L^2 \alpha_2}{u^2}, \quad f_3^2 = L^2 \alpha_4  \nonumber \\
4 \rho^2 \left(dr^2 + r^2 d\theta^2\right) = \frac{4L^2du^2}{u^2} + L^2 \alpha_3 d\tilde{\theta}^2 \label{eq:FGfunctionmap2}
\end{gather}
We regulate the spacetime at a small value of $u$ and identify this with $\varepsilon$: the (dimensionless) UV cut-off on $AdS_3\times S^3$. The boundary conditions on the coordinate map and the functions $\alpha_i(u,\tilde\theta)$ at small $u$ must be chosen to ensure that the boundary metric is $ds^2_{AdS_3} +ds^2_{S^3}$ and the transverse $S^4$ is recovered.  We find
\begin{gather}
r = {2 \over  u^2} +\ldots, \quad \theta = \tilde\theta +\ldots 
\nonumber \\
\alpha_1 = 1 + \ldots, \quad \alpha_2 = 1+ \ldots, \quad \alpha_3 = 1 +\ldots,  \quad \alpha_4 = \sin^2\tilde\theta + \ldots  \label{eq:alphaBCs2}
\end{gather}
Whilst we have not been able to solve \eqref{eq:FGfunctionmap2} in closed form, we can build the coordinate map as an asymptotic expansion in $u$.  The mapping is given by
\begin{align}
r &= {2 \over  u^2} + {m_2 \cos \tilde\theta\over 4 } +{3(-16- m_2^2 + 8 m_3) + (16 -21 m_2^2+ 40  m_3) \cos2\tilde\theta \over 768 }\, u^2\nonumber \\
&\phantom{=\ } +\frac{\cos \tilde\theta}{18432 }\left( 48  m_2-43 m_2^3 + 40  m_2 m_3 + 80 m_4\right.\nonumber\\
&\phantom{=\ }\left. -\left( 48 m_2 -203 m_2^3+680  m_2 m_3 - 560 m_4\right)\cos2\tilde\theta\right)  u^4 +O(u^6)\label{rthuya}
\end{align}
and
\begin{align}
\theta&=\tilde\theta -{m_2\sin \tilde\theta\over 8 }\, u^2 -{(16 -27 m_2^2 + 40  m_3)\cos \tilde\theta \sin \tilde\theta \over 768  }\,  u^4 \nonumber \\
&\phantom{=\ }+\frac{\sin \tilde\theta}{18432 } \left(296 m_2 m_3 -48 m_2 -98 m_2^3 -200 m_4\right.\nonumber \\
&\phantom{=\ }\left. + \left(48 m_2-139 m_2^3 + 400  m_2 m_3 -280  m_4\right)\cos 2\tilde\theta\right) u^6 + O(u^8) \label{rthuyb}
\end{align}

The area integral and action integral both diverge at large $r$.  It is useful to express the coordinate map as a cut-off relation $r_c=r_c(\theta,\varepsilon)$.  This is found by first inverting the relation (\ref{rthuyb}) in the small $u$ limit and then eliminating $\tilde\theta$ from (\ref{rthuya}). The result is
\begin{align}
r_c(\theta,\varepsilon)&={2 \over  \varepsilon^2} +{m_2 \cos\theta\over 4 } + {-3(16 + 5m_2^2- 8 m_3) + (16 -9 m_2^2+40 m_3) \cos 2\theta \over 768 }\, \varepsilon^2 \nonumber\\
&\phantom{=\ }+\frac{\cos\theta}{9216}
 \left(-48  m_2 + 55 m_2^3 -    160  m_2 m_3 +    40  m_4 \right. \nonumber \\
 &\phantom{=\ }\left. +\left(48  m_2 + 25 m_2^3 - 
      160 m_2 m_3 + 280  m_4\right) \cos 2 \theta\right) \varepsilon^4+ O(\varepsilon^6) \label{rcutrel}
\end{align}

\section{Holographic entanglement entropy}
\setcounter{equation}{0}
\label{sec3}

The Ryu-Takayanagi prescription \cite{Ryu:2006bv,Ryu:2006ef} states that the entanglement entropy of  a spatial region ${\cal A}$  is given by the area of a co-dimension two  minimal surface $\cal M$ in the bulk that is anchored on the $AdS$ boundary at $\partial {\cal A}$:
\begin{align}\label{eq:EE1}
S_{\cal A}= {A_{\mathrm{min}} \over 4 G_N^{(11)}}
\end{align}
Since we are dealing with static states of our CFT, this surface lies on a constant time slice. If this surface is not unique, we choose the one whose area is minimal among all such surfaces homologous to ${\cal A}$.\footnote{This minimal surface prescription  was recently established on a firm footing by the analysis of \cite{Lewkowycz:2013nqa}.}

In the following section we derive the minimal surface $\cal M$ for a general bubbling solution and show that its restriction to the boundary maps to a four-sphere in $\mathbb{R}^6$.  We then evaluate its regulated area and compare with our expectations from $\mathbb{R}^6$.

\subsection{Minimal surface geometry}

A bubbling geometry is an $AdS_3\times S^3\times S^3$ fibration over $\Sigma$.  We consider a surface ${\cal M}$ at constant $t$ that fills the $S^3\times S^3$ and has profile $z=z(w,\bar{w},l)$, where $z$ is the $AdS_3$ radial coordinate defined in \eqref{eq:AdS3metric}. The area functional becomes 
\begin{align}\label{eq:areaintegral}
A({\cal M}) &=  2\, \textrm{Vol}(S^3)^2 \int dl \int_{\Sigma} d^2w \, {f_1 f_2^3 f_3^3 \rho^2 \over z}\,   \sqrt{1+ {f_1^2\over  z^2\rho^2} {\partial z\over\partial w}  {\partial z\over  \partial \bar w} + \left( {\partial z\over \partial l}\right)^2}
\end{align}
The equations of motion derived from this functional are solved by
\begin{align}\label{eq:surface}
z(w,\bar{w},l)^2 + l^2 &= R^2
\end{align}
This semicircle is simply a co-dimension two minimal surface in $AdS_3$. Following  \cite{Jensen:2013lxa,Estes:2014hka}  it is straightforward to see that this is in fact the surface of  minimal area (within this ansatz).

The surface \eqref{eq:surface} is independent of the $AdS_7$ radial coordinate.  Thus,  the boundary $\partial {\cal A}$ of the entangling region on $AdS_3\times S^3$ is given by the same formula.   To understand this, let us consider two coordinate charts on $\mathbb{R}^6$:
\begin{align}
\label{eq:Weylrescaling}
ds^2_{\mathbb{R}^6}&= z^2 \left( \frac{dz^2+dt^2+dl^2}{z^2} + ds_{S^3}^2\right) = dt^2 + d\bar{r}^2 +\bar{r}^2 \left(d\chi^2 +\sin^2\chi\,  ds^2_{S^3} \right)
\end{align}
The map between these two charts is given by
\begin{align}
z&=\bar{r} \, \sin\chi,\quad\quad l = \bar{r}\, \cos\chi
\end{align}
Thus, our $\partial {\cal A}$ on $AdS_3\times S^3$ can be written as a four-sphere of radius $R$ on $\mathbb{R}^6$ (given by $\bar{r}=R$) after a Weyl rescaling.

\subsection{Evaluating the area integral}\label{evalarea}

The combination of metric factors that appears in the area integral \eqref{eq:areaintegral} can be written 
\begin{align}
f_1 f_2^3 f_3^3 \rho^2 & = \frac{1}{4}\, |\partial_w h|^2\, h\, W_-
\end{align} 
The entanglement entropy is proportional to the area evaluated on the surface \eqref{eq:surface}: 
\begin{align}\label{eq:EE2}
S_{\cal A}&=  \frac{\textrm{Vol}(S^3)^2}{8 G_N^{(11)}} \int dl\, \frac{R}{R^2-l^2}\, (J_1+J_2)
\end{align}
where we have defined 
\begin{align}
J_1 &\equiv \int_\Sigma d^2w\,  |\partial_w h|^2 \; h \;  |G-\bar{G}| \nonumber \\
J_2 &\equiv -2 \int_\Sigma d^2w\,  |\partial_w h|^2 \; h \;  |G|^2
\end{align}
Substituting $w = r\, e^{i\theta}$ into \eqref{gdefa} we find for $J_1$
\begin{align}
J_1 = - 4 L^9 \sum_{i=1}^{2 n} (-1)^i \int_0^{\pi} d\theta\, \sin\theta \int_0^{r_c\left(\theta,\varepsilon\right)} dr\, \frac{r^2 (r\cos\theta - \xi_i)}{\sqrt{r^2+\xi_i^2-2r \xi_i \cos\theta}} \label{eq:J1definition}
\end{align}
The overall minus sign follows from the fact that $G-\bar{G} <0$ on the upper half-plane.  
We carefully evaluate this expression in appendix~\ref{jonecalc}.  The final result  is given in equation  \eqref{appjone} and takes the form
\begin{align}
J_1 &= L^9 \left[\frac{64 }{3  \varepsilon^4} + \frac{-24+3m_2^2-8m_3}{15} + O\left(\varepsilon^2\right) \right]
\end{align}
Next we consider the second term 
 \begin{align}
J_2 &= -2 \int_\Sigma d^2w\,  |\partial_w h|^2 \; h \;  |G|^2 \nonumber \\
 & = - 2  L^9 \int_0^{\pi} d\theta\, \sin\theta \int_0^{r_c\left(\theta,\varepsilon\right)} dr\,  r^2 \nonumber \\
    &\phantom{=\ } \times\left\{ 2 n + 2 \sum_{i<j} (-1)^{i+j} \frac{r^2-r\cos\theta\, (\xi_i+\xi_j)+\xi_i\xi_j}{\sqrt{r^2-2 r \xi_i \cos\theta+\xi_i^2} \sqrt{r^2-2 r \xi_j \cos\theta+\xi_j^2}} \right\} \label{eq:J2expanded}
\end{align}
We carefully evaluate this integral in appendix~\ref{appjtwo} and the final result is \eqref{eq:J2final}
\begin{align}
J_2 &= L^9 \left[-\frac{64 }{3\varepsilon^2} - \frac{4}{3} \sum_{i<j} (-1)^{i+j} |\xi_i-\xi_j|^3+O\left(\varepsilon^2\right) \right]
\end{align}
Note that the second term cannot be expressed in terms of the moments $m_k$. 

Now we handle the integral over $l$.  Recall that the minimal surface formula \eqref{eq:surface} describes a semicircle for which $z\in[0,R]$ and $l\in[-R,R]$.  Note that $J_{1,2}$ are independent of $l$ because the cut-off function is.  The $l$  integral diverges at both limits; rewriting via \eqref{eq:surface}  as an integral over $z$, we regulate with a cut-off at $z=\eta$:
\begin{align}\label{AdS3cut}
\int_{-\sqrt{R^2-\eta^2}}^{\sqrt{R^2-\eta^2}} dl\, \frac{R}{R^2-l^2}&=2 \int_0^{\sqrt{R^2-\eta^2}} dl\, \frac{R}{R^2-l^2} = 2  \int_{\eta}^R dz\, \frac{R}{z \sqrt{R^2-z^2}} \nonumber\\
 &= 2 \log\left(\frac{R + \sqrt{R^2 - \eta^2}}{\eta}\right) = 2\log\left(\frac{2R}{\eta}\right) - \frac{\eta^2}{2 R^2} + O(\eta^4)
\end{align}

Finally we put these pieces together to compute the divergent entanglement entropy \eqref{eq:EE2}:
\begin{align}
S_{\cal A}&=  \frac{ L^9 \,\textrm{Vol}(S^3)^2}{4 G_N^{(11)}} \, \left[ \frac{64 }{3\varepsilon^4}  -\frac{64 }{3\varepsilon^2} + \frac{-24+3m_2^2-8m_3}{15}\right. \nonumber \\
&\phantom{=\ }\left.- \frac{4}{3} \sum_{i<j} (-1)^{i+j} |\xi_i-\xi_j|^3 +O\left(\varepsilon^2\right)  \right] \log\left(\frac{2R }{\eta}\right)
\end{align}
Employing the definitions 
\begin{align}
\label{eq:somedefinitions}
L&= (\pi\, N)^{1/3}\, \ell_P, \quad \quad 8\pi G_N^{(11)} = 2^7 \pi^8 \ell_P^9, \quad \quad \textrm{Vol}(S^3) = 2\pi^2
\end{align}
this becomes
\begin{align}
S_{\cal A}&=  \frac{ 4 N^3}{3} \, \left[ \frac{1 }{\varepsilon^4}  -\frac{1 }{\varepsilon^2} + \frac{-24+3m_2^2-8m_3}{320}\right. \nonumber \\
&\phantom{=\ }\left.- \frac{1}{16} \sum_{i<j} (-1)^{i+j} |\xi_i-\xi_j|^3  +O\left(\varepsilon^2\right) \right] \log\left(\frac{2R }{\eta}\right) \label{eq:EEfinalunsubtracted}
\end{align}
Evaluating this result on the vacuum we find
\begin{align}
\label{eq:EEvacuum}
S_{\cal A}^{(0)}&=\frac{ 4 N^3}{3} \, \left[  \frac{1 }{\varepsilon^4}  -\frac{1 }{\varepsilon^2} + \frac{3}{8}
  +O\left(\varepsilon^2\right) \right] \log\left(\frac{2R }{\eta}\right)
\end{align}
Subtracting this vacuum contribution from \eqref{eq:EEfinalunsubtracted} we arrive at our final result for the change  in  entanglement entropy due to the presence of the Wilson surface:
\begin{align}
\Delta S_{\cal A}&=\frac{ 4 N^3}{3} \, \left[  \frac{16+3  m_2^2-8 m_3}{320}
 - \frac{1}{16 } \sum_{i<j} (-1)^{i+j} |\xi_i-\xi_j|^3 -\frac{1}{2}\right] \log\left(\frac{2R }{\eta}\right)\label{eq:EEfinal}
\end{align}
Note that the power divergences with respect to the FG cut-off $\varepsilon$ are cancelled in this subtraction and only a logarithmic divergence in $\eta$ remains.

\subsection{Physical interpretation}
\label{sec33}

In this section we give a physical interpretation of our result for the entanglement entropy. First, recall that during the calculation we introduced two separate regulators. The FG cut-off $\varepsilon$ can be viewed as a regular UV cut-off for a holographic theory with  a six-dimensional  $AdS_3\times S^3$ boundary. In addition, when performing the integral over the $AdS_3$ coordinate $l$ in (\ref{AdS3cut}) we  introduced a cut-off $\eta$ on the $AdS_3$ radial coordinate $z$ in Poincare slicing (\ref{eq:AdS3metric}).
One might be tempted to  view $\eta$ as purely an IR cut-off that regulates the infinite volume of the boundary theory. However, it  also has an interpretation as a UV cut-off  on the minimal distance to the Wilson surface in the boundary theory. 

This interpretation is most easily demonstrated by considering the vacuum spacetime.  The map that relates  $AdS_7\times S^4 $ with    $AdS_3\times S^3$ boundary in (\ref{adsmeta})  to a metric with $\mathbb{R}^6$ boundary
\begin{align}\label{eq:PAdS7}
ds^2&= \frac{4L^2}{\tilde u^2} \left( d \tilde u^2 + dt^2 +dl^2+ dr^2 +r^2  ds_{S^3}^2  \right) + L^2  \left( dq^2 + \sin^2 q\, ds^2_{\tilde S^3}\right)
\end{align}
is given by 
\begin{align}
\label{eq:cmap}
z &= \tilde u\cosh p    \quad \quad r = \tilde u\sinh p  
\end{align}
Setting  $\tilde u=\tilde \varepsilon$ imposes a (dimensionful) holographic UV cut-off on the theory living on the $\mathbb{R}^6$ boundary. 
Since the minimal value of the coordinate $p$ is zero, it follows from (\ref{eq:cmap})  that the range of $z$ is bounded by $z>\tilde \varepsilon$ and hence the $AdS_3$ cut-off $\eta$ is related to the uniform UV cut-off $\tilde \varepsilon$.

 At this point we do not have an analog of the map (\ref{eq:cmap})  for the bubbling solution that is valid for all values of the coordinates. We can construct the map for the asymptotic region defined by  the FG expansion for the $AdS_3\times S^3$  boundary theory (\ref{eq:FG}), however this expansion breaks down  once the FG coordinate $u$ is not small. As discussed in \cite{Estes:2014hka} one can construct a map that is valid also near $z=0$ by patching together the expansion near the $AdS_3$ boundary and the FG boundary. 
 We will not pursue this construction here since we focus all our calculations on the theory living on the $AdS_3\times S^3$ boundary. However, since the metric is asymptotically $AdS$, we expect that one  should obtain only a small modification  to the identification of the UV cut-offs that, crucially,   does not affect the logarithmically divergent term in the entanglement entropy.

Physically, the interpretation of $\eta$ as a UV cut-off and the form of the subtracted entanglement entropy \eqref{eq:EEfinal} is quite natural. Note that  the  dominant contributions to the entanglement entropy come from UV degrees of freedom located near the entangling surface. Since   the Wilson surface (at fixed time) intersects the entangling surface at two points in our geometry (see figure \ref{fig:entanglesurf}), the defect contribution to the entanglement entropy has essentially the same dimensionality as the entanglement entropy of a two-dimensional CFT. This  is also reflected in the $AdS_3$ slicing we employ and the fact that the minimal surface we find in the bulk \eqref{eq:surface}  is familiar from $AdS_3$/CFT$_2$. Hence an argument along the lines of those given in  \cite{Estes:2014hka}  shows  that the extra divergent contribution of the Wilson surface should be logarithmic, wherein the cut-off is associated with a    minimal distance to the defect.

\section{Holographic   stress tensor}
\setcounter{equation}{0}
\label{sec4}

The goal of the present section is to calculate the one point function of the six-dimensional stress tensor holographically for an asymptotically $AdS_7\times S^4$ bubbling solution. Since the bubbling solutions are eleven-dimensional one has to utilize the machinery of KK holography that was developed in \cite{Skenderis:2006uy}. Note that a similar calculation was performed in \cite{Gomis:2008qa} for the type IIB bubbling solution dual to half-BPS Wilson loop defects and we will largely adopt their method to our case.

The KK reduction of  eleven-dimensional supergravity  on $S^4$  produces a seven-dimensional supergravity (with negative cosmological constant) with infinite towers of massive fields \cite{Pilch:1984xy,vanNieuwenhuizen:1984iz}. These can be classified by their seven-dimensional spin and  representation  of the relevant $SO(5)$ spherical harmonics (scalar and tensorial) on $S^4$. One difficulty is the mixing of modes coming from     the eleven-dimensional supergravity as well as the fact that seven-dimensional fields can be related by eleven-dimensional diffeomorphisms, leading to nonlinear gauge symmetries. The resulting seven-dimensional action can be diagonalized and the masses of all the fields were determined in   \cite{Pilch:1984xy,vanNieuwenhuizen:1984iz,Gunaydin:1984wc}.

Via the AdS/CFT correspondence the seven-dimensional supergravity fields are dual to operators in the six-dimensional  $(2,0)$-theory. The precise dictionary  can be found for example in  
\cite{Leigh:1998kt}. 

In  \cite{Skenderis:2006uy} it was argued that in order to obtain a local seven-dimensional supergravity action without higher derivatives one needs in general to perform a KK reduction map that is nonlinear and relates the eleven- and seven-dimensionall fields schematically as
\begin{align}
\Psi_{7}&=\psi_{11}+ {\cal K}\, \psi_{11}\psi_{11}+\ldots
\end{align}
Here, ${\cal K}$ is a differential operator and the ellipsis denotes higher order terms with three or more eleven-dimensional  fields. This nonlinear mixing in general  complicates holographic calculations (see e.g.\ \cite{Gomis:2008qa}). 
  However, a simple rule was derived  in \cite{Skenderis:2006uy} to determine when and which nonlinear terms appear.  For a supergravity field dual to a dimension $\Delta$ operator in the CFT, the only nonlinear terms  that can appear  are the ones for which the sum of the  dimensions of their respective  dual operators is less than or equal to $\Delta$. When this rule is applied to the stress tensor, which has $\Delta=6$, it is clear that there can be no nonlinear mixing since the  operators with lowest dimension have $\Delta=4$.\footnote{There is a `doubleton' field dual to an operator of dimension $\Delta=2$, but such fields are free and decouple from the dynamics.} 
  
  The starting point for the calculation of the holographic stress tensor is to decompose the eleven-dimensional metric into a $AdS_7\times S^4$ part and a perturbation, denoted as $g^{(0)}$ and $h$ respectively:
\begin{align}
\label{metadSpert}
ds^2&= g_{M N}\, dx^M dx^N = \left( g_{M N}^{(0)} + h_{M N} \right) dx^M dx^N.
\end{align}
We use the FG coordinate chart (\ref{eq:FG}) where we can identify $\tilde{\theta}$ as the polar angle on $S^4$. The Wilson surface preserves an $SO(4)$ subgroup of the R-symmetry and so does the bubbling geometry. Therefore, performing the harmonic decomposition on the eleven-dimensional fields we obtain contributions only from spherical harmonics invariant under $SO(4)$. These depend only on the polar angle $\tilde{\theta}$. The zero mode on $S^4$ of an eleven-dimensional field that only has nontrivial dependence on $\tilde{\theta}$  can be expressed as
\begin{align}
\bar \phi(x) &= {\int_0^\pi  d\tilde{\theta} \; \phi(x,\tilde{\theta}) \sin^3\tilde{\theta} \over \int_0^\pi  d\tilde{\theta} \sin^3\tilde{\theta} }
\end{align}
The reduced seven-dimensional metric, which satisfies the seven-dimensional linearized Einstein equation, is given by the following combination\footnote{Our index conventions are: $M, N,\ldots$ are eleven-dimensional indices, $\mu,\nu, \ldots$ are $AdS_7$ indices and $a,b, \ldots$ are $S^4$ indices.} 
\begin{align}
\label{sevenmet}
ds_7^2&= \left[\left(1+ {1\over 5} \bar \pi \right)  g^{(0)}_{\mu\nu}   + \bar h_{\mu\nu}\right] dx^\mu dx^\nu
\end{align}
Here, $ g^{(0)}$ is the $AdS_7$ vacuum metric whereas  $\bar h_{\mu\nu}$ is the zero mode of the fluctuations in the seven dimensions. The field $ \bar \pi $ is the zero mode of the trace of the fluctuations of the metric along the $S^4$ directions:
\begin{align}
\pi(x,\tilde\theta)&\equiv   h^{a b}\, g^{(0)}_{ab}
\end{align}
The factor of ${1\over 5}$ in (\ref{sevenmet}) comes from a Weyl rescaling to bring the   KK reduced metric to the Einstein frame in seven dimensions. 

The calculation of the one point function of the stress tensor from the seven-dimensional metric (\ref{sevenmet}) utilizes  the standard method of holographic renormalization. In our case $\bar{\pi}$ vanishes, and consequently the reduced metric is already in the FG form
\begin{align}
ds_7^2&= \frac{4 L^2}{u^2}\left( du^2 +  g_{ij} dx^i dx^j\right)
\end{align}
where the large $r$ limit corresponds to $u\to 0$ and the  metric $g_{ij}$ can be expressed as a power series in $u$.
The holographic stress tensor can then be calculated immediately using the formulae  for $d=6$ given in   \cite{deHaro:2000xn}. For the convenience of the reader and completeness we present the details of these calculations in appendix~\ref{appstress}. We have also checked that the counter-term approach developed in \cite{Kraus:1999di} gives the same result for the stress tensor.  The final result for the change in the expectation value of the stress tensor in the presence of the Wilson surface on $AdS_3\times S^3$ is 
 \begin{align}
\Delta \langle T_{i j} \rangle\, dx^{i} dx^{j}& = \frac{N^3}{160 \pi^3} \left(16 + 3m_2^2 - 8m_3\right)   \left( ds^2_{AdS_3} - ds^2_{S^3} \right) 
\end{align}

Note that the dependence of $N^3$ is as expected from a back-reacted  supergravity solution. In addition, the expression depends only on  the first two nontrivial moments $m_2,m_3$ of the bubbling solution.
The stress tensor contribution has a form that respects the $so(2,2|\mathbb{R})\oplus so(4|\mathbb{R})$ of the Wilson surface and is traceless in line with the absence of a conformal anomaly on $AdS_3\times S^3$.
The use of this result is two-fold. First, any nontrivial holographic observable is useful to understand $(2,0)$ theory
better. Second, the stress tensor expectation value is an important ingredient in the calculation of the entanglement entropy using the replica trick, as we discuss briefly in \ref{sec6}.

\section{Expectation value of the Wilson surface operator}
\setcounter{equation}{0}
\label{sec5}

The expectation value for the Wilson surface operator can be obtained from the following formula:
\begin{align}
\label{eq:WSformula}
\langle W_{\Gamma}\rangle &= \exp\left[-(I-I_{(0)})\right]
\end{align}
where $I$ is the eleven-dimensional supergravity action evaluated on a general  bubbling solution and $I_{(0)}$ is the  action evaluated on the $AdS_7\times S^4$ vacuum. The action is given by
\begin{align}
I&={1\over 8 \pi G_N^{(11)}}  \left[ \frac{1}{2}\int d^{11}x\, \sqrt{g}\left( R-{1\over 48}F_{MNPQ}F^{MNPQ}\right) - \frac{1}{12} \int C\wedge F\wedge F \right. \nonumber \\
 &\phantom{=\ }\left. +   \int d^{10} x\,  \sqrt{\gamma}\, K\right]\label{sugraaction}
\end{align}
The final term is the Gibbons-Hawking boundary term, which is necessary to make the variational principle well-defined for spacetimes with boundary. 
The metric functions for the bubbling solution are given in \eqref{met2} and the four-form field strength is given by
\begin{align}
F&= (f_1)^3 g_{1m} \, \omega_{AdS_3}\wedge e^m + (f_2)^3 g_{2m}  \, \omega_{S^3}\wedge e^m + (f_3)^3g_{3m} \, \omega_{\tilde S^3}\wedge e^m 
\end{align}
where $e^m$ are the vielbeins on the Riemann surface $\Sigma$,  $\omega_X$ denotes the volume form for a unit-radius space $X$ and the expressions for $g_{I m}$ with $I=1,2,3$ can be found in appendix~\ref{appast}.

\subsection{Action as a total derivative}

\begin{figure}[!t]
  \centering
  \includegraphics[width=90mm]{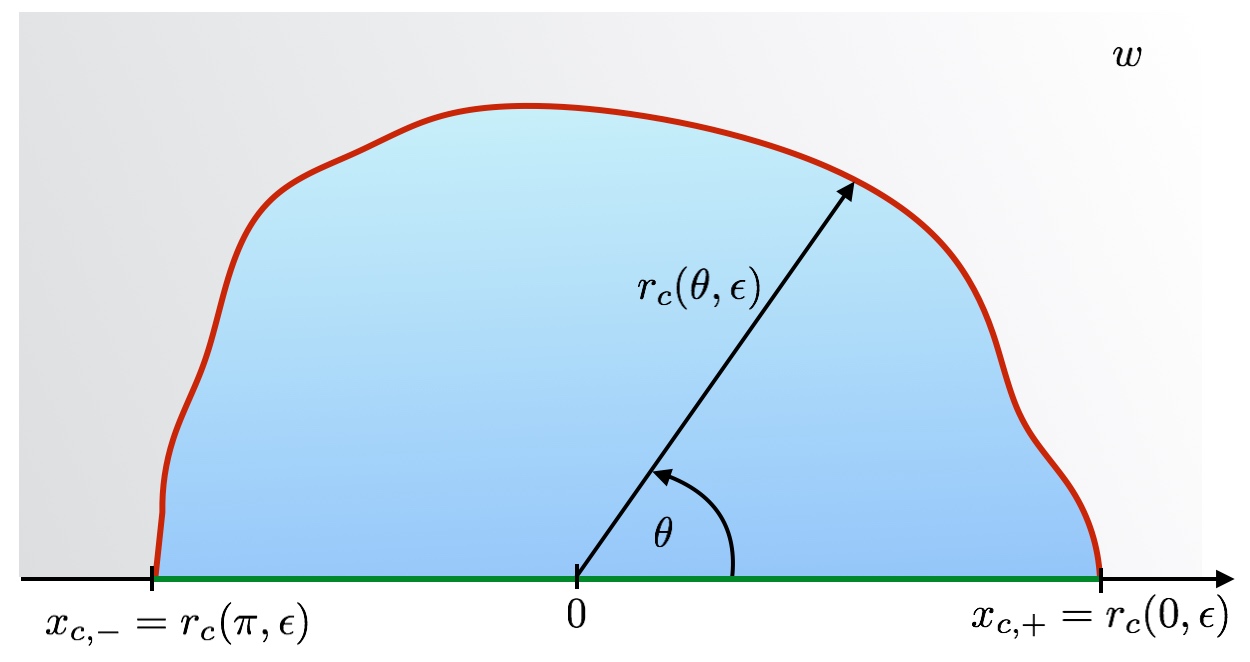}
  \caption{The two boundary components: the cut-off surface (red) and the boundary along $x\equiv \Re w$ (green). }
  \label{fig:vacuumgeometryb}
\end{figure}

First we demonstrate the well-known fact that the on-shell action of eleven-dimensional supergravity is a total derivative. Using the Einstein equation, the Ricci scalar can be eliminated from the bulk term of the on-shell supergravity action (\ref{sugraaction}):
\begin{align}
I_{\rm bulk} ={1\over 16 \pi G_N^{(11)}} \int \Big( -{1\over 3} F\wedge * F -{1\over 6} C\wedge F\wedge F\Big)
\label{actnor}
\end{align}
The equation of motion can be expressed as
\begin{align}
d * F + {1\over 2} F\wedge F& =0
\end{align}
The first term in the action can then be written as
\begin{align}
F\wedge * F  &= dC \wedge *F  \nonumber \\
&=  d(C  \wedge *F) + C \wedge d* F \nonumber \\
&= d(C  \wedge *F)-{1\over 2} C\wedge F\wedge F
\end{align}
Plugging this expression  into (\ref{actnor}) the $C\wedge F\wedge F$ terms cancel and the on-shell value of the action is indeed a total derivative:
\begin{align}
\label{actexp1}
I_{\rm bulk} &= -{1\over 48 \pi G_N^{(11)}} \int d(C  \wedge * d C) 
\end{align}
The duals of the three contributions to the  field strength are given by
\begin{align}
\label{dualfex}
*F_{(1)}&= {f_2^3f_3^3\over  f_1^3}   \Big(r\partial_r b_1 d\theta -{1\over r} \partial_\theta b_1 dr\Big)  \wedge \omega_{S^3}\wedge \omega_{\tilde S^3}\nonumber\\
*F_{(2)}&= {f_1^3f_3^3\over  f_2^3}   \Big(r\partial_r b_2 d\theta -{1\over r} \partial_\theta b_2 dr\Big)  \wedge \omega_{AdS_3}\wedge \omega_{\tilde S_3}\nonumber\\
*F_{(3)}&= {f_1^3f_2^3\over  f_3^3}   \Big(-r\partial_r b_3 d\theta +{1\over r} \partial_\theta b_3 dr\Big)  \wedge \omega_{AdS_3}\wedge \omega_{ S^3}
\end{align}
Using the expression for the metric functions  $f_i$ from   (\ref{met2}), the ratios   appearing in (\ref{dualfex}) can be expressed as
 \begin{align}
 \label{metrfaca}
 \left({f_2 f_3\over f_1}\right)^3&= {h W_-^2\over 4  W_+(1-|G|^2)} \nonumber\\
\left({f_1 f_3\over f_2}\right)^3&= {h W_+^2\over 4  W_-(1-|G|^2)} \nonumber\\
\left({f_1 f_2\over f_3}\right)^3&= {16 h (1-|G|^2)^2\over W_+ W_-}  
 \end{align}
 Since the volume forms on the $AdS_3$ and the three-spheres are all closed, the bulk action (\ref{actexp1}) reduces to a total derivative over the two-dimensional Riemann surface $\Sigma$.  In terms of the polar coordinates  $r,\theta$, which were introduced in section~\ref{sec:asymreg} by setting $w=r e^{i \theta}$,
the bulk part of the action can be written as follows:
\begin{align}
  I_{\rm bulk}&=  -{1\over 48 \pi G_N^{(11)}}\int d(C\wedge *F \nonumber) \nonumber\\
  &= -{1\over  48 \pi G_N^{(11)}}\int \left( \partial_r a_r + \partial_\theta a_\theta\right) dr\wedge d\theta \wedge  \omega_{AdS_3} \wedge\omega_{S^3}\wedge\omega_{\tilde S^3} \label{eq:bulkactionintermediate}
\end{align}
  where $a_r$ and $a_\theta$ are given by
\begin{align}
\label{arathdef}
a_r&= -{f_2^3f_3^3\over 2 f_1^3} r\partial_r (b_1^2)+{f_1^3f_3^3\over 2 f_2^3} r\partial_r (b_2^2)+{f_1^3f_2^3\over 2 f_3^3} r\partial_r (b_3^2)\nonumber\\
a_\theta &= -{f_2^3f_3^3\over 2 f_1^3}{1\over  r}\partial_\theta (b_1^2)+{f_1^3f_3^3\over 2 f_2^3} {1\over  r}\partial_\theta(b_2^2)+{f_1^3f_2^3\over 2 f_3^3}{1\over  r}\partial_\theta (b_3^2)
\end{align}
We have shown that the on-shell action can be written as a boundary term:   the bulk term reduces to a total derivative and the  Gibbons-Hawking term is also boundary term.  The boundary in the $w$ plane has two pieces (see figure~\ref{fig:vacuumgeometryb}). First one has the cut-off surface parametrized by  $r_c(\theta,\varepsilon)$, where $\varepsilon$ is the FG cut-off. Second one has the real line, which is parametrized in $r,\theta$ coordinates by $r\in [0, r_c (0,\pi,\varepsilon)]$ and $\theta=0, \pi$, respectively. In the following  we will evaluate the contribution from each piece in turn. 

\subsection{Gibbons-Hawking term}
\label{ghterma}
The Gibbons-Hawking term in (\ref{sugraaction}) is the boundary term that has to be added to the action in order to give a well-defined  gravitational variational principle.  A ten-dimensional surface  defined by a constraint
\begin{align}
\label{constr}
F\left(x^M\right)&=0
\end{align}
can be parameterized by a set of coordinates $\sigma^{\alpha}$.  The induced metric is defined via
\begin{align}
  \gamma_{\alpha\beta}&=e^M_\alpha\, e^N_\beta\,  g_{MN}, \quad e^M_\alpha = {\partial x^M\over \partial \sigma^\alpha}
\end{align}
and its determinant is given by
\begin{align}
\sqrt{\gamma} &=  {1\over z^3}\,  2 \rho  f_1^3 f_2^3 f_3^3 \, \omega_{S^3} \, \omega_{\tilde S^3} 
 \end{align}
The normal  vector (which we assume to be always space-like) is defined via
\begin{align}
\label{kabform1}
\hat n_M &= {1\over  \sqrt{ {\partial F\over \partial x^N} {\partial F\over \partial x^P} g^{NP} }}\,  {\partial F\over \partial x^M}
\end{align}
The extrinsic curvature  and its trace are  defined via
\begin{align}
\label{kabform2}
K_{\alpha\beta} &= \left( \nabla_M \hat n_N  \right) e_\alpha^M\,  e_\beta^N,\quad   K= \gamma^{\alpha\beta} K_{\alpha\beta}
\end{align}
As discussed above the boundary has two components, which we now study in turn.

\subsubsection{Real line contribution}

The real line is defined by $F=y=0$, where $w=x+i\, y$, and the induced metric is simply obtained by dropping  the $g_{yy}$ component  of the eleven-dimensional metric. The normal vector is given by
\begin{align}
\hat n^M &= {1\over 2 \rho}\, \delta^M_y
\end{align}
Using the form of the metric it is straightforward to  determine the trace of the extrinsic curvature
\begin{align}
K&=  {1\over L^3} \left({3 \partial_y   f_1 \over 2 \rho f_1 }+ {3 \partial_y   f_2 \over 2 \rho f_2 }+{3 \partial_y   f_3 \over 2 \rho f_3 }+ {\partial_ y \rho \over 2 \rho^2} \right)
\end{align}
 The Gibbons-Hawking term along the real line can be determined  from the expansion of the metric functions in the $y\to 0$ limit given.  This  can be obtained from the formulae in appendix \ref{apprealx}. We do not present the details of this calculation but just present the final result which is 
  \begin{align}
  \int d^{10} x\,  \sqrt{\gamma}\, K& =   \lim_{y\to 0} \textrm{Vol}(S^3)^2\, \textrm{Vol}(AdS_3)\, L^9 \int dx  \left[ 24  y^2 + O(y^3) \right] =0
 \end{align}
 which vanishes. This result was expected  since at any point on the real line, one of the three-spheres shrinks to zero size and the space closes off.

\subsubsection{Large $r$ contribution}
In this section we determine the contribution of the Gibbons-Hawking term from the large $r$ cut-off surface defined by the equation
\begin{align}
F(r,\theta)&= r-r_{c}(\theta,\varepsilon)=0
\end{align}
 for small $\varepsilon$, where $r_c(\theta,\varepsilon)$ is given  by (\ref{rcutrel}). Hence the surface extends along $AdS_3$ and the two three-spheres and the induced metric in these directions  is identical to the metric. We choose to use $\theta$ to parametrize the cut-off surface and find the following nontrivial components of the vielbein $e^M_\alpha$:
 \begin{align}
e^\alpha_{\beta} &=\delta^{\alpha}_\beta, \quad e^r_\theta = \partial_\theta\,  r_c(\theta,\varepsilon)
\end{align}
Using the formulae (\ref{kabform1}) and (\ref{kabform2}) we can calculate $\sqrt{\gamma}\, K$ and expand the result in a power series in $\varepsilon$:
\begin{align}
\sqrt{\gamma}\, K &=L^9 \sin^3 \theta \Big\{ \frac{192}{\varepsilon^6} + \frac{96 m_2 \cos\theta}{\varepsilon^4}  \nonumber \\
&\phantom{=\ } \left. + \frac{-3(16 -3m_2^2-40m_3) + 5 \cos{2\theta} (16+3m_2^2+40m_3)}{8 \varepsilon^2}\right. \nonumber \\
&\phantom{=\ } \left.   \frac{1}{4} \cos\theta \left[ -8 m_2 + m_2^3-6 m_2 m_3 +10  m_4 \right.\right. \nonumber \\
&\phantom{=\ } \left.  + \cos {2\theta} ( 24 m_2+m_2^3-10 m_2 m_3 +70  m_4) \right]            \Big\} + O\left(\varepsilon^2\right)
\end{align}
After performing the integral over $\theta$ we find that the Gibbons-Hawking term is determined completely by the    contribution at large $r$, namely
\begin{align}
\label{actionGH}
I_{\textrm{GH}}& \equiv  {1\over 8 \pi G_N^{(11)}}\, \int d^{10} x\,  \sqrt{\gamma}\, K
= {L^9\over 8 \pi G_N^{(11)}}\, \textrm{Vol}(S^3)^2\, \textrm{Vol}(AdS_3) \left[ {256\over  \varepsilon^6} -{16\over \varepsilon^2} + O\left(\varepsilon^2\right)\right]
\end{align}

\subsection{Bulk supergravity action}

The bulk part of the eleven-dimensional action integral \eqref{eq:bulkactionintermediate} can be reduced to the following form:
 \begin{align}
 \label{actinta}
I&=  -{1\over 48 \pi G_N^{(11)}}\,  \textrm{Vol}(S^3)^2\, \textrm{Vol}(AdS_3) \int d\theta \int dr \left(\partial_r a_r + \partial_\theta a_\theta\right) \nonumber \\
&=   -{1\over 48 \pi G_N^{(11)}}\, \textrm{Vol}(S^3)^2\, \textrm{Vol}(AdS_3) \oint_{\partial \Sigma(\varepsilon)} d\tau\, (\hat n_r a_r+ \hat n_\theta a_\theta) 
  \end{align}
 Here $\textrm{Vol}(AdS_3)$ is the (regularized) volume of  $AdS_3$ since we consider the field theory on $AdS_3\times S^3$ and  $\hat n_i$ is the unit outward normal.  If a boundary component is parametrized by $r(\tau), \theta(\tau)$ then the unit outward  normal vector is defined as
 \begin{align}
 \hat n_r &= {d\theta\over d\tau}\, d\tau  , \quad \quad \hat n_\theta =-  {dr \over d\tau}\, d\tau
 \end{align}
 There are three components to the boundary:
 \begin{align}
 \partial \Sigma_1:&\  \theta\in[0,\pi],\; r= r_c( \theta,\varepsilon)  \nonumber\\ 
   \partial \Sigma_2:&\  \theta=0 , r\in [0, r_c(0, \varepsilon) ]\nonumber \\
     \partial \Sigma_3:&\    \theta=\pi,\; r\in [0, r_c(\pi , \varepsilon)] 
 \end{align}
 The contribution of $\partial \Sigma_1$ corresponds to the cut-off surface, which  we parametrize  by $\theta\in[0,\pi]$ and  $r=r_c(\theta, \varepsilon)$ as in section \ref{ghterma}. The $\theta$ dependence of $r$ leads to an additional contribution to the normal vector 
 \begin{align}
 \label{lrcuoff}
 I_{\rm (cut)} &=  -{1\over  48 \pi G_N^{(11)}} \,  \textrm{Vol}(S^3)^2\, \textrm{Vol}(AdS_3)\oint_{\partial \Sigma_1(\varepsilon)} d\tau\, (\hat n_r a_r+ \hat n_\theta a_\theta)  \nonumber \\
 &=   -{1\over  48 \pi G_N^{(11)}} \,  \textrm{Vol}(S^3)^2\, \textrm{Vol}(AdS_3)\int_0^\pi d\theta \left.\left( a_r-{\partial r_c\over\partial \theta}\, a_\theta\right)\right|_{r=r_c(\theta,\varepsilon)}
 \end{align}
The  integrations over $\partial \Sigma_2$ and $ \partial \Sigma_3$ combine to give the integration over the real line. Putting the two together and going back to the half-plane coordinates we find
 \begin{align}
  \oint_{\partial \Sigma_2(\varepsilon)+\partial \Sigma_3(\varepsilon) } d\tau \, (\hat n_r a_r+ \hat n_\theta a_\theta)
&=  - \int_0^{r_c(\theta=0,\varepsilon)} dr \left. {1\over r}\, a_\theta \right|_{\theta=0}
+ \int_0^{r_c(\theta=\pi,\varepsilon)} dr \left. {1\over r}\, a_\theta \right|_{\theta=\pi}
  \end{align}
  Using ${1\over r} \partial_\theta|_{\theta=0} = {\partial_y}$ and $ {1\over r} \partial_\theta |_{\theta=\pi } =-\partial_y$ and the expressions for $a_r$ and $a_\theta$ given in (\ref{arathdef}), the contribution from the real line becomes 
  \begin{align}
  \label{actionxb}
  I_{(x)}&= {1\over  48 \pi G_N^{(11)}} \,  \textrm{Vol}(S^3)^2\, \textrm{Vol}(AdS_3) \int_{r_c(\theta=\pi,\varepsilon)}^{r_c(\theta=0,\varepsilon)}dx 
  \left(- {f_2^3 f_3^3\over 2 f_1^3} \partial_y (b_1^2) +{f_1^3 f_3^3\over 2 f_2^3} \partial_y (b_2^2)+{f_1^3 f_2^3\over 2 f_3^3} \partial_y (b_3^2)\right)
  \end{align}

 \subsubsection{Large $r$ contribution}
 Using the expansion of  $a_r$ and $a_\theta$ at large $r$ one finds the integrand of (\ref{lrcuoff}) can be written as
 \begin{align}
  \frac{1}{L^9} \left.\left( a_r-{\partial r_c\over\partial \theta} a_\theta\right) \right|_{r=r_c(\theta,\varepsilon)}&= \frac{5 \sin\theta \cos^2\theta (\cos{2\theta}-5) (16+3m_2^2-8 m_3)}{4 \varepsilon^2} \nonumber \\
&\phantom{=\ } + {\sin{2\theta}\over {128  }} \left[(2144 m_2+445m_2^3-1760 m_2 m_3 -860 m_4  \right.\nonumber \\
&\phantom{=\ }+ 20 \cos{2\theta} \left(64 m_2+ 15m_2^3-80 m_2 m_3-60 m_4 \right)\nonumber \\
&\phantom{=\ } \left.+\cos {4\theta} \left(-96 m_2-25m_2^3+160 m_2 m_3+140 m_4 \right)\right]\nonumber \\
&\phantom{=\ } +O\left(\varepsilon^2\right)
 \end{align}
 After performing the integration over $\theta$ the finite terms above  drop out and  and we are left with
 \begin{align}
 \label{actionAtCutoff}
  I_{\rm (cut)} &= -{L^9 \over  48 \pi G_N^{(11)}}\,  \textrm{Vol}(S^3)^2\, \textrm{Vol}(AdS_3) \left[\frac{-64 -12 m_2^2+ 32 m_3}{\varepsilon^2} +O\left(\varepsilon^2\right) \right]
 \end{align}
 
 \subsubsection{Real line contribution}
 \label{sec532}

The expression from the real line can be obtained from expanding the integrand of (\ref{actionxb}) in the $y\to 0$ limit.  The behavior of the integrand in this limit depends on the interval  $x$ is located in.  We define
 \begin{align}
  \cI_0&= [ -\infty,\xi_1] \cup [\xi_2,\xi_3] \cup \cdots \cup   [\xi_{2n},+\infty] \nonumber\\
   \cI_+&= [ \xi_1,\xi_2] \cup [\xi_3,\xi_4] \cup \cdots \cup   [\xi_{2n-1},\xi_{2n}] 
    \end{align}
    We present the details of the calculation in appendix \ref{apprealx}, with the final result  given by
    \begin{align}
    \label{ylimint}
     \lim_{y\to 0}\left(- {f_2^3 f_3^3\over 2 f_1^3} \partial_y (b_1^2) +{f_1^3 f_3^3\over 2 f_2^3} \partial_y (b_2^2)+{f_1^3 f_2^3\over 2 f_3^3} \partial_y (b_3^2)\right)=\left\{\begin{array}{lc}
{16 L^9 (2 g_1^3+ g_3) \phi_0 \over (g_1^2-g_2)(g_1^2+g_2)}& x\in \cI_0  \\
 \\
 {32 L^9 (g_1^3+ 3 g_1 g_2-g_3)(\phi_0+ 2x)\over (g_1^2+g_2)(g_1^2+ 2 g_2)} &
x\in \cI_+ 
\end{array}
\right.
    \end{align}
where the  $g_i$ and $\phi_0$ are functions of $x$ and can be found in (\ref{gidefa}) and (\ref{phi0defa}), respectively. In this limit the integrand (\ref{ylimint}) is nonsingular  for any finite $x$ on the real line. However, the integrand grows as $x\to \pm \infty$ and we can extract the divergent behavior coming from the region near the cut-off. We determine the divergent contributions in (\ref{apsxdiv}) and we write the  result as 
\begin{align}
\label{actionAtRealLine}
I_{(x)}&= + {L^9\over  48 \pi G_N^{(11)}}\,  \textrm{Vol}(S^3)^2\, \textrm{Vol}(AdS_3) \left[ -\frac{256}{\varepsilon^6}+ \frac{80-12m_2^2+32m_3}{\varepsilon^2}+  {\cal F} +O\left(\varepsilon^2\right) \right] 
\end{align}
The finite term ${\cal F}$ can in principle be determined by performing the $x$-integral for the full integration region and subtracting the divergent contributions given above, i.e.\ 
\begin{align}
\label{finitepart}
{\cal F}  & \equiv \lim_{\varepsilon\to 0} \left\{ \sum_{x\in \cI_+} \int dx\,  {32 (g_1^3+ 3 g_1 g_2-g_3)(\phi_0+ 2x)\over (g_1^2+g_2)(g_1^2+ 2 g_2)} 
 +\sum_{x\in \cI_0} \int dx \,  {16(2 g_1^3+ g_3) \phi_0 \over (g_1^2-g_2)(g_1^2+g_2)}\right.\nonumber \\
 &\phantom{=\ }\left. - \left(  -\frac{256}{\varepsilon^6}  +\frac{80-12m_2^2+32m_3}{\varepsilon^2} \right)  \right\}
\end{align}
The last two terms remove the divergent contributions from the integral over $\cI_0$ that is regulated for large positive $x$ by   $x<x_{c,+} (\varepsilon)$ and for large  negative $x$ by $x>x_{c,-} (\varepsilon) $. 

At this point we have not been able to find a closed expression for these integrals, but they can be evaluated numerically. If a relation to a matrix model calculation exists these integrals may  be related to the resolvent. However, as no proposal for a matrix model exists at present  we have not pursued the evaluation further and leave this for future work.

\subsection{Final result}
Combining all the contributions to the action, which can be found in \eqref{actionGH}, \eqref{actionAtCutoff} and \eqref{actionAtRealLine}, we obtain for the on-shell action 
\begin{align}
I  &= \frac{L^9}{3 \pi G_N^{(11)}}\,  \textrm{Vol}(S^3)^2\, \textrm{Vol}(AdS_3) \left[ \frac{80}{\varepsilon^6} + \frac{3}{\varepsilon^2}+\frac{\cal F}{16}+O\left(\varepsilon^2\right) \right] \nonumber \\
 &= \frac{N^3}{12 \pi}\, \textrm{Vol}(AdS_3) \left( \frac{80}{\varepsilon^6} + \frac{3}{\varepsilon^2}+\frac{\cal F}{16}+O\left(\varepsilon^2\right)\right)
\end{align}
The finite contribution  is given by (\ref{finitepart}).  This can be evaluated exactly for the vacuum and the result is
\begin{align}
{\cal F}_{(0)}& = -64
\end{align}
Thus, using \eqref{eq:WSformula} we can express our final result for the expectation value of the Wilson surface operator as 
\begin{align}
\log\, \langle W_{\Gamma} \rangle &= - \frac{N^3}{192 \pi}\, \textrm{Vol}(AdS_3) \left({\cal F}+64\right)
\end{align}
 Note that the power divergences in $\varepsilon$ are   independent of the details of the bubbling geometry and so cancel in the subtraction. However, the result is proportional to the infinite volume of $AdS_3$, which is regulated by the cut-off $z=\eta$ introduced in the entanglement entropy calculation in (\ref{AdS3cut}).

As is the case for the stress tensor contribution, the expectation value of the Wilson surface operator constitutes a potentially useful holographic observable of the $(2,0)$ theory.  This holographic result  may be compared to direct calculations in this theory as well as be applied to the replica calculation of the entanglement entropy.  It would be very interesting to study localization and related methods to calculate the expectation value of  Wilson surface operators  in the future.

\section{Discussion}
\setcounter{equation}{0}
\label{sec6}

In this  paper we have calculated the holographic entanglement entropy for the six-dimensional $(2,0)$ theory in the presence of a Wilson surface. In addition we have calculated two other holographic observables:  the one point function of the stress tensor in the presence of a Wilson surface and  the expectation value of the Wilson surface operator.  The bubbling solution that provides the holographic dual of the `heavy' Wilson surface is determined in terms of $2n$ real numbers $\xi_i$ that parametrize the partition of the real line into two types of segment (where $G=0$ or $G=+i$). The quantities we calculated all depend on moments $m_k$  whose definition in terms of $\xi_i$  is given in (\ref{momentdefa}). In particular, the stress tensor only depends on $m_2$ and $m_3$.  On the other hand,    the final term in the  result for the entanglement entropy (\ref{eq:EEfinal}) cannot in general  be expressed in terms of the moments $m_k$.  The finite part of the expectation value of the Wilson surface operator is expressed in terms of a real integral.  While it is possible to evaluate this finite part for $n=1$ and $n=2$ in closed form, we have been unable to evaluate it in general. Some numerical experiments however indicate that this integral does not have a simple expression in terms of the moments $m_k$ or the final term in the entanglement entropy (\ref{eq:EEfinal}).

  This `coloring' of the real line is reminiscent of other bubbling solutions, such as those dual to  Wilson loops \cite{D'Hoker:2007fq} or  the LLM solutions \cite{Lin:2004nb}, where the data of the solution can be mapped to the data of a matrix model or free fermion phase space, respectively. One hint that BPS Wilson surfaces might be described by matrix models can be seen as follows. If one compactifies the six-dimensional $(2,0)$ theory on a  circle one obtains five-dimensional SYM theory \cite{Douglas:2010iu,Minahan:2013jwa} and a Wilson surface that wraps the circle becomes a Wilson line in this five-dimensional gauge theory \cite{Young:2011aa,Mori:2014tca}. Various quantities have been calculated for this system using localization and it would be very interesting to see if these results can be lifted to six dimensions and compared to our calculations in the limit of very large `representations', in analogy with \cite{Gentle:2014lva}.
  
  It would also be interesting  to see  whether  the calculation of the entanglement entropy in the presence of a Wilson loop in  $SU(N)$  ${\cal N}=4$ SYM  due to  Lewkowycz and Maldacena  \cite{Lewkowycz:2013laa} generalizes   to the Wilson surface in the $(2,0)$ theory.  Recall that their calculation used  the replica trick and involved  the expectation value of the Wilson loop and the stress tensor in the presence of the Wilson loop on the space $S^1\times H^3$.  A generalization would most likely start from the expectation value the Wilson surface  and the stress tensor in the presence of the Wilson surface on $S^1\times H^5$.   Since our holographic calculation gives these two quantities for the $AdS_3\times S^3$ boundary, if it is possible to map the results to  $S^1\times H^5$ then it should be possible to compare the Lewkowycz and Maldacena calculation to the holographic entanglement entropy calculation we  have performed.   One complication is that unlike a one-dimensional Wilson loop, the two-dimensional Wilson surface  has a conformal anomaly \cite{Graham:1999pm,Henningson:1999xi,Gustavsson:2004gj} and it is not clear how to determine its contribution in our case. A simpler case in which to consider the anomaly  might be the case of an abelian Wilson surface, as studied in \cite{Gustavsson:2004gj}.
  
 Our calculations provide  results that could  be compared to any  field theory or localization calculation in the $(2,0)$ theory. What is missing at this point, compared to the analogous case of the half-BPS Wilson loop in ${\cal N}=4$ SYM, is a concrete dictionary between a matrix model and the supergravity parameters as well as a better understanding of the anomaly and the  map of the stress tensor and the expectation value to the $S^1\times H^5$ boundary geometry. We leave these interesting questions for future work.

\section*{Acknowledgements}

It is a pleasure to thank Eric D'Hoker, Per Kraus and Christoph Uhlemann for useful discussions.  This work was supported in part by National Science Foundation grant PHY-13-13986.

\newpage

\appendix 

\section{ Contributions to the entanglement entropy}
\label{app:extrafinitepiece}
In this appendix we carefully discuss the contribution  to the area integrals that are needed in section \ref{evalarea}.

\subsection{$J_1$}\label{jonecalc}
First we consider $J_1$, given in \eqref{eq:J1definition}. The radial integral can be performed directly, but it is useful to rewrite it in terms of Legendre polynomials using \eqref{eq:Legendre}.  We divide the integration range into two regions: $0\leq r \leq |\xi_i|$ and $|\xi_i|\leq r \leq r_c\left(\theta,\varepsilon\right)$.  For each region we choose the  Legendre representation that converges, yielding
 \begin{align}
J_1 &= -4 L^9 \sum_{i=1}^{2 n} (-1)^i \int_0^{\pi} d\theta\, \sin\theta 
 \left\{\int_{|\xi_i|}^{r_c\left(\theta,\varepsilon\right)} dr\, \frac{r^2 (r\cos\theta - \xi_i)}{r} \sum_{\ell=0}^\infty P_\ell\,(\cos\theta) \left(\frac{\xi_i}{r}\right)^\ell \right. \nonumber \\
&\phantom{=\ } \left. +\int_0^{|\xi_i|} dr\, \frac{r^2 (r\cos\theta - \xi_i)}{|\xi_i|} \sum_{\ell=0}^\infty P_\ell\,(\cos\theta) \left(\frac{r}{\xi_i}\right)^\ell  \right\}
\end{align}
Performing the two radial integrals directly we find
 \begin{align}
J_1 &= -4 L^9 \sum_{i=1}^{2 n} (-1)^i \int_0^{\pi} d\theta\, \sin\theta 
\left\{ \left[  \cos\theta\, P_0\, \frac{r^3}{3} + (\cos\theta\, P_1 - P_0)\, \xi_i\, \frac{r^2}{2} + 
  (\cos\theta\, P_2 - P_1)\, \xi_i^2\, r  \right. \right. \nonumber \\
   &\phantom{=\ } \left.  + (\cos\theta\, P_3 - P_2)\, \xi_i^3\, \log r - \sum_{\ell=1}^\infty \frac{(\cos\theta P_{\ell+3}-P_{\ell+2})}{\ell}\, \frac{\xi_i^{\ell+3}}{r^{\ell}}\right]_{|\xi_i|}^{r_c\left(\theta,\varepsilon\right)} \nonumber \\
   &\phantom{=\ } + \left.  \frac{1}{|\xi_i|}\left[ - P_0\, \xi_i\, \frac{r^3}{3} +\sum_{\ell=4}^\infty \frac{(\cos\theta P_{\ell-4}-P_{\ell-3})}{\ell}\, \frac{r^{\ell}}{\xi_i^{\ell-4}} \right]_0^{|\xi_i|} \right\}
\end{align}
Orthogonality of the Legendre polynomials can be expressed via
\begin{align}
\label{eq:Legendreorthogonality}
\int_0^{\pi} d\theta\, \sin\theta  P_\ell\,(\cos\theta) P_k\,(\cos\theta) &= \frac{2}{2 \ell + 1} \, \delta_{\ell k}
\end{align}
We use this to simplify the above expression dramatically:
 \begin{align}
J_1 &= -4 L^9 \sum_{i=1}^{2 n} (-1)^i   \left\{ \int_0^{\pi} d\theta\, \sin\theta \left[  \cos\theta\, P_0\, \frac{r_c^3}{3} + (\cos\theta\, P_1 - P_0)\, \xi_i\, \frac{r_c^2}{2} + 
  (\cos\theta\, P_2 - P_1)\, \xi_i^2\, r_c  \right. \right. \nonumber \\
   &\phantom{=\ } \left.  \left.+ (\cos\theta\, P_3 - P_2)\, \xi_i^3\, \log r_c - \sum_{\ell=1}^\infty \frac{(\cos\theta P_{\ell+3}-P_{\ell+2})}{\ell}\, \frac{\xi_i^{\ell+3}}{r_c^{\ell}}\right] +\frac{2}{15}\, \xi_i^3 \right\} \label{eq:J1almost}
\end{align}
Note that the final term is a sum of contributions at $r=|\xi_i|$.  Substituting  for the   cut-off function $r_c(\theta,\varepsilon)$ given in \eqref{rcutrel}, we then expand in $\varepsilon$ up to and including $O\left(\varepsilon^0\right)$ and perform the remaining integrals over $\theta$. We find the final result
\begin{align}\label{appjone}
J_1 &= L^9 \left[\frac{64 }{3  \varepsilon^4} + \frac{-24+3m_2^2-8m_3}{15} + O\left(\varepsilon^2\right) \right]
\end{align}

\subsection{$J_2$}\label{appjtwo}

Next we calculate $J_2$, which we reproduce from \eqref{eq:J2expanded}:
 \begin{align}
J_2  & = - 2  L^9 \int_0^{\pi} d\theta\, \sin\theta \int_0^{r_c\left(\theta,\varepsilon\right)} dr\,  r^2 \nonumber \\
    &\phantom{=\ } \times\left\{ 2 n + 2 \sum_{i<j} (-1)^{i+j} \frac{r^2-r\cos\theta\, (\xi_i+\xi_j)+\xi_i\xi_j}{\sqrt{r^2-2 r \xi_i \cos\theta+\xi_i^2} \sqrt{r^2-2 r \xi_j \cos\theta+\xi_j^2}} \right\} \label{eq:J2expandedapp}
\end{align}
We can split the integral into two terms coming from the sum in the last line  of \eqref{eq:J2expandedapp}.
The first term is simply
\begin{align} 
J_{2,a}&\equiv -\frac{4}{3}\, n L^9  \int_0^{\pi} d\theta\, \sin\theta\, r_c\left(\theta,\varepsilon\right)^3 \nonumber\\
&= L^9 \left[ -\frac{64 n}{3 \varepsilon^6}+ \frac{n (40  + 3 m_2^2 - 8 m_3)}{18 \varepsilon^2} +O\left(\varepsilon^2\right) \right] \label{eq:J2a}
\end{align}

The evaluation of the second term, denoted $J_{2,b}\equiv J_2 - J_{2,a}$, is more involved than that of $J_1$. Our strategy is to divide up the radial integration range and replace the square root factors with the appropriate  convergent series of Legendre polynomials in each interval.  The fraction in the summand is symmetric under $(i\leftrightarrow j)$ so we can choose $|\xi_i|<|\xi_j|$ without loss of generality and write: 
 \begin{align}
J_{2,b} &= -4 L^9 \sum_{i<j} (-1)^{i+j} \int_0^{\pi} d\theta\, \sin\theta \nonumber \\
&\phantom{=\ } \times \left\{\int_{|\xi_j|}^{r_c\left(\theta,\varepsilon\right)} dr\, \frac{r^2 (r^2-r\cos\theta\, (\xi_i+\xi_j)+\xi_i\xi_j)}{r^2} \sum_{\ell=0}^\infty P_\ell\left(\frac{\xi_i}{r}\right)^\ell \sum_{k=0}^\infty P_k \left(\frac{\xi_j}{r}\right)^k \right. \nonumber \\
 &\phantom{=\ } +\int_{|\xi_i|}^{|\xi_j|} dr\, \frac{r^2 (r^2-r\cos\theta\, (\xi_i+\xi_j)+\xi_i\xi_j)}{r |\xi_j|} \sum_{\ell=0}^\infty P_\ell \left(\frac{\xi_i}{r}\right)^\ell \sum_{k=0}^\infty P_k \left(\frac{r}{\xi_j}\right)^k \nonumber \\
&\phantom{=\ } \left. +\int_0^{|\xi_i|} dr\, \frac{r^2 (r^2-r\cos\theta\, (\xi_i+\xi_j)+\xi_i\xi_j)}{|\xi_i| |\xi_j|} \sum_{\ell=0}^\infty P_\ell\left(\frac{r}{\xi_i}\right)^\ell \sum_{k=0}^\infty P_k \left(\frac{r}{\xi_j}\right)^k   \right\} \nonumber  \\
&\equiv  K_1 + K_2 + K_3  \label{eq:J2bseparate}
\end{align}
where the Legendre polynomials are all functions of $\cos\theta$, as before.  
First let us consider $K_1$: 
 \begin{align}
K_1 &\equiv -4 L^9 \sum_{i<j} (-1)^{i+j}  \int_0^{\pi} d\theta\, \sin\theta   \nonumber \\
&\phantom{=\ } \times \sum_{\ell,k=0}^\infty P_\ell  P_k\, \xi_i^\ell \xi_j^k \int_{|\xi_j|}^{r_c\left(\theta,\varepsilon\right)} dr\, \left(r^2-r\cos\theta\, (\xi_i+\xi_j)+\xi_i\xi_j\right) r^{-\ell-k}  
\end{align}
We must perform the radial integral first because its upper limit depends on $\theta$.  This results in several sums over powers of $r$ and a logarithm:  
 \begin{align}
K_1  &= -4 L^9 \sum_{i<j} (-1)^{i+j}  \int_0^{\pi} d\theta\, \sin\theta    \left\{ \sum_{\substack{\ell,k=0 \\
       \ell+k\neq 3}}^\infty \frac{P_\ell  P_k\, \xi_i^\ell \xi_j^k\, r^{3-\ell-k}}{3-\ell-k}\right. \nonumber \\ 
&\phantom{=\ } -\cos\theta(\xi_i+\xi_j) \sum_{\substack{\ell,k=0 \\
       \ell+k\neq 2}}^\infty \frac{P_\ell  P_k\, \xi_i^\ell \xi_j^k\, r^{2-\ell-k}}{2-\ell-k}
 +\xi_i \xi_j \sum_{\substack{\ell,k=0 \\
       \ell+k\neq 1}}^\infty \frac{P_\ell  P_k\, \xi_i^\ell \xi_j^k\, r^{1-\ell-k}}{1-\ell-k} \nonumber \\
 &\phantom{=\ } \left. + \left( \sum_{\substack{\ell,k=0 \\
       \ell+k= 3}}^3 P_\ell  P_k\, \xi_i^\ell \xi_j^k - \cos\theta(\xi_i+\xi_j)\sum_{\substack{\ell,k=0 \\
       \ell+k= 2}}^2 P_\ell  P_k\, \xi_i^\ell \xi_j^k + \xi_i \xi_j \sum_{\substack{\ell,k=0 \\
       \ell+k= 1}}^1 P_\ell  P_k\, \xi_i^\ell \xi_j^k \right)\log r \right\}_{|\xi_j|}^{r_c\left(\theta,\varepsilon\right)} \label{eq:K1full}
\end{align}
We only require the entanglement entropy up to and including $O\left(\varepsilon^0\right)$. Recall that the  cut-off function $r_c\left(\theta,\varepsilon\right)$ given  \eqref{rcutrel} leads with $O\left(\varepsilon^{-2}\right)$, and therefore only the logarithm and non-negative powers of $r$   contribute to the upper limit.  Specifically, we can terminate the infinite sums in the first and second lines at $3, 2$ and $1$, respectively.  Integrating over $\theta$ we find
\begin{align}
\label{eq:J2bdivpart}
K_1^{\textrm{upper}}&= L^9 \left[\frac{64 n }{3 \varepsilon^6}- \frac{n (40  + 3 m_2^2 - 8 m_3)}{18 \varepsilon^2} -\frac{64}{3 \varepsilon^2}+O\left(\varepsilon^2\right) \right]
\end{align}
where we have made use of the following results:
\begin{align}
\sum_{i<j} (-1)^{i+j} &= - n, \quad \quad \sum_{i<j} (-1)^{i+j} (\xi_i-\xi_j)^2=- 4
\end{align}
Next let us consider the lower limit and perform the integral over $\theta$. The terms with no explicit $\cos\theta$ factor vanish unless $\ell = k$ by orthogonality  \eqref{eq:Legendreorthogonality}.  To deal with the terms that do have an explicit $\cos\theta$ factor, let us  define
\begin{align}
X_{\ell k}&\equiv \int_0^{\pi} d\theta\, \sin\theta \cos\theta P_\ell\,(\cos\theta) P_k\,(\cos\theta) = 2 \left( \begin{array}{ c c c } 1 & \ell & k \\ 0 & 0 & 0 \end{array} \right)^2  \label{eq:Legendreorthogonality2} \\
&= \frac{2(\ell-k)^2 (1+\ell+k)}{(\ell+k)(2+\ell +k)(1+\ell-k)!(1-\ell+k)!}\nonumber
\end{align}
These terms are only non-zero when $X_{\ell k}$ is too, which occurs when $|\ell - k| = 1$.  These two observations imply that the coefficient of  the logarithm vanishes and  that the conditions on the sums in the first two lines of \eqref{eq:K1full} have no effect for the lower limit.  All that remains is
 \begin{align}
K_1^{\textrm{lower}}  &= +4 L^9 \sum_{i<j} (-1)^{i+j}  \sum_{\ell,k=0}^\infty \xi_i^{\ell}\xi_j^k \left[\frac{2  }{2 \ell + 1} \, \delta_{\ell k} \left(\frac{|\xi_j|^{3-\ell-k}}{3-\ell-k} + \frac{|\xi_j|^{1-\ell-k}}{1-\ell-k}\, \xi_i\xi_j\right) \right. \nonumber \\
 &\phantom{=\ } \left. - \frac{|\xi_j|^{2-\ell-k}}{2-\ell-k }\, (\xi_i+\xi_j)\, X_{\ell k}\right]
\end{align}
Performing the sum over $k$ and using the definition \eqref{eq:Legendreorthogonality2} we find
 \begin{align}
K_1^{\textrm{lower}}  &= +4  L^9\sum_{i<j} (-1)^{i+j}  \sum_{\ell=0}^\infty \frac{2  }{2 \ell + 1}\, \xi_i^{\ell}\xi_j^{\ell} \left[ \frac{|\xi_j|^{3-2\ell}}{3-2\ell} +\frac{|\xi_j|^{1-2\ell}}{1-2\ell} \, \xi_i\xi_j \right. \nonumber \\
 &\phantom{=\ } \left.- \frac{(\ell+1)\, |\xi_j|^{1-2\ell}}{(2\ell+3)(1-2\ell)} \, (\xi_i+\xi_j)^2\right]
\end{align}
The limits on the radial integrals in $K_{2,3}$ are independent of $\theta$  so we are free to reverse the order of integration.  Let us begin with $K_2$:
\begin{align}
K_2 &\equiv -4 L^9 \sum_{i<j} \frac{(-1)^{i+j}}{|\xi_j|}  \int_{|\xi_i|}^{|\xi_j|} dr\, \sum_{\ell,k=0}^\infty r^{k-\ell}\, \frac{\xi_i^{\ell}}{\xi_j^k} \nonumber \\ 
&\phantom{=\ }
 \times \int_0^{\pi} d\theta\, \sin\theta P_\ell  P_k\left( r^3-r^2\cos\theta\, (\xi_i+\xi_j)+r\xi_i\xi_j\right) \nonumber \\
 &= -4  L^9 \sum_{i<j} \frac{(-1)^{i+j}}{|\xi_j|}  \int_{|\xi_i|}^{|\xi_j|} dr\, \sum_{\ell,k=0}^\infty r^{k-\ell}\, \frac{\xi_i^{\ell}}{\xi_j^k} \left[\frac{2}{2 \ell + 1} \, \delta_{\ell k}\, (r^3 + r \xi_i\xi_j)- r^2 (\xi_i+\xi_j) X_{\ell k}\right] 
\end{align}
where again we have used \eqref{eq:Legendreorthogonality} and \eqref{eq:Legendreorthogonality2}.  Performing the sum over $k$ then integrating over $r$ we find
\begin{align}
K_2 &= -4 L^9 \sum_{i<j} \frac{(-1)^{i+j}}{|\xi_j|}   \sum_{\ell=0}^\infty \frac{2}{2\ell+1}\, \frac{\xi_i^{\ell}}{\xi_j^{\ell}}\left[1 - \frac{\ell+1}{2\ell+3}\left( 1+\frac{\xi_i}{\xi_j} \right)\right]   \nonumber \\ 
&\phantom{=\ }
 \times \left( \frac{|\xi_j|^4-|\xi_i|^4}{4} +\frac{|\xi_j|^2-|\xi_i|^2}{2} \, \xi_i\xi_j   \right) 
 \end{align}
We can compute $K_3$ using the same method:
\begin{align}
K_3 &\equiv -4 L^9 \sum_{i<j} \frac{(-1)^{i+j}}{|\xi_i||\xi_j|}  \int_{0}^{|\xi_i|} dr\, \sum_{\ell,k=0}^\infty  \frac{r^{\ell+k}}{\xi_i^{\ell}\xi_j^k} \nonumber \\ 
&\phantom{=\ }
 \times \int_0^{\pi} d\theta\, \sin\theta P_\ell  P_k\left( r^4-r^3\cos\theta\, (\xi_i+\xi_j)+r^2\xi_i\xi_j\right) \nonumber \\
 &= -4 L^9 \sum_{i<j} \frac{(-1)^{i+j}}{|\xi_i||\xi_j|}  \int_{0}^{|\xi_i|} dr\, \sum_{\ell,k=0}^\infty  \frac{r^{\ell+k}}{\xi_i^{\ell}\xi_j^k} \left[\frac{2}{2 \ell + 1} \, \delta_{\ell k}\, (r^4 + r^2 \xi_i\xi_j)- r^3 (\xi_i+\xi_j) X_{\ell k}\right] \nonumber \\
 &= -4 L^9 \sum_{i<j} \frac{(-1)^{i+j}}{|\xi_i||\xi_j|}   \sum_{\ell=0}^\infty \frac{2}{2\ell+1}\, \frac{1}{\xi_i^{\ell}\xi_j^{\ell}}\left\{\left[1 - \frac{\ell+1}{2\ell+3}\left( \xi_i +\xi_j \right) \left( \frac{1}{\xi_i} +\frac{1}{\xi_j} \right)\right] \frac{|\xi_i|^{5+2\ell}}{5-2\ell} \right. \nonumber \\ 
&\phantom{=\ }
 + \left. \frac{|\xi_i|^{3+3\ell}}{3+2\ell}  \, \xi_i\xi_j   \right\} 
\end{align}

Now we combine the finite contributions to $J_{2,b}$.  The infinite sums can be evaluated and the following remarkably simple result is obtained:
\begin{align}
K_1^{\textrm{lower}} +K_2+K_3 &= \frac{4 L^9}{3} \sum_{i<j} (-1)^{i+j} \sgn  \xi_j \left(\xi_i-\xi_j\right)^3
\end{align}
Recall that we have assumed $|\xi_i|<|\xi_j|$.  The ordering of the $\xi_i$ in \eqref{eq:xiordering} implies that $\sgn  \xi_j$ evaluates to $+1$  and since the sum is ordered  we can write the finite contribution to $J_{2,b}$ as 
\begin{align}
-\frac{4 L^9}{3} \sum_{i<j} (-1)^{i+j} |\xi_i-\xi_j|^3
\end{align}
Note that this term cannot be expressed in terms of the moments $m_k$. Thus, our final result for $J_{2,b}$ is:
\begin{align}\label{jtwoappres}
J_{2,b}&=L^9 \left[\frac{64 n }{3 \varepsilon^6}- \frac{n (40  + 3 m_2^2- 8 m_3)}{18 \varepsilon^2} -\frac{64}{3 \varepsilon^2} -\frac{4}{3} \sum_{i<j} (-1)^{i+j} |\xi_i-\xi_j|^3 +O\left(\varepsilon^2\right) \right]
\end{align}

Summing  \eqref{eq:J2a} and \eqref{jtwoappres} we find
\begin{align}\label{eq:J2final}
J_2 &= L^9 \left[-\frac{64 }{3\varepsilon^2} - \frac{4}{3} \sum_{i<j} (-1)^{i+j} |\xi_i-\xi_j|^3+O\left(\varepsilon^2\right) \right]
\end{align}

\section{Calculation of the holographic stress tensor}
\setcounter{equation}{0}
\label{appstress}
In this appendix we present some details of the KK reduction calculation as well as the   calculation of the stress tensor using holographic renormalization. As mentioned in section~\ref{sec4}, first one has to decompose the metric into the vacuum $AdS_7 \times S^4$ part and fluctuations, as in (\ref{metadSpert}). In  FG coordinates the vacuum metric is given by
\begin{align}
\label{FGAdS}
g_{MN}^{(0)}\, dx^M dx^N &= L^2 \left[\frac{4}{u^2}\left(du^2+ \left(1+\frac{u^2}{2}+\frac{u^4}{16} \right) ds^2_{AdS_3} +  \left(1-\frac{u^2}{2}+\frac{u^4}{16} \right) ds^2_{S^3} \right)   \right. \nonumber \\
&\phantom{=\ } \left. + d\tilde{\theta}^2 + \sin^2 \tilde{\theta} ds^2_{\tilde{S}_3} \right] .
\end{align}
The metric fluctuations in terms of the functions $\alpha_i (u,\tilde{\theta})$ appearing in (\ref{eq:FG}) are
\begin{align}
h_{MN}\, dx^M dx^N &=  L^2 \left[\frac{4}{u^2} \left\{ \left( \alpha_1 -1-\frac{u^2}{2}-\frac{u^4}{16} \right) ds^2_{AdS_3} +\left( \alpha_2 - 1+\frac{u^2}{2}-\frac{u^4}{16} \right)  ds^2_{S^3} \right\}  \right. \nonumber \\
&\phantom{=\ } \left. + \left(\alpha_3-1\right) d\tilde{\theta}^2 + \left(\alpha_4-\sin^2 \tilde{\theta} \right) ds^2_{\tilde{S}_3} \right].
\end{align}
Using these expressions, we calculate the seven-dimensional reduced metric (\ref{sevenmet}) and the outcome is
\begin{align}
\label{reducedmet}
ds^2_7 &= \frac{4L^2}{u^2} \left[ du^2 + \left( 1 + \frac{u^2}{2} +\frac{u^4}{16} +\frac{1}{320} (16+ 3 m_2^2- 8 m_3) u^6 \right) ds^2_{AdS_3}\right. \nonumber \\
&\phantom{=\ } \left. + \left(1 - \frac{u^2}{2} +\frac{u^4}{16} -\frac{1}{320} (16+ 3 m_2^2- 8 m_3) u^6 \right) ds^2_{S^3}  \right]
\end{align}
Notice that substituting the vacuum moments in (\ref{reducedmet}) one can retrieve the $AdS_7$ entries in (\ref{FGAdS}). This is because the trace shift does not contribute to the reduced metric, i.e.\ $\bar{\pi}$ vanishes. Furthermore, a further FG map of (\ref{reducedmet}) is not necessary since it is already in FG form:
\begin{align}
\label{fgappa}
ds_7^2&= \frac{4 L^2}{u^2}\left( du^2 +  g_{ij}\, dx^i dx^j\right)
\end{align}
where the six-dimensional metric $g_{ij}$ is given by a power series in $u$:
 \begin{align}
 \label{fgappb}
g&=g_{(0)}+  g_{(2)}\, u^2+ g_{(4)}\, u^4 + g_{(6)}\,  u^6+ h_{(6)}\, u^6 \log u^2+\ldots
\end{align}

To compute the holographic stress tensor, we simply read off the asymptotic metric coefficients $g_{(0)}$, $g_{(2)}$, $g_{(4)}$ and $g_{(6)}$ from \eqref{reducedmet} and substitute them into the $d=6$ formula   given in  \cite{deHaro:2000xn}. For completeness we present this fomula here:
\begin{align}
\langle T_{ij}\rangle& ={3\, (2L)^5 \over 8\pi G_N^{(7)}} \left(g_{(6) ij}- A_{(6)ij}+{1\over 24}\, S_{ij}\right)
\end{align}
where the second and third terms are defined via
\begin{align}
A_{(6) ij}&= {1\over 3}\left[ 2 (g_{(2)} g_{(4)})_{ij} + (g_{(4)}g_{(2)})_{ij} - (g_{(3)}^3)_{ij} +{1\over 8}g_{(2) ij} \left( \tr g_{(2)}^2 -(\tr g_{(2)})^2\right)\right.\nonumber\\
&\phantom{=\ }- \tr g_{(2)} \left( g_{(4)ij} -{1\over 2} (g_{(2)}^2)_{ij} \right)- g_{(0)ij}\left( {1\over 8} \tr g_{(2)}^2 \tr  g_{(2)}-{1\over 24}(\tr g_{(2)})^3 \right.\nonumber\\
&\phantom{=\ }\left.\left.-{1\over 6} \tr g_{(2)}^3 +{1\over 2} \tr(g_{(2)} g_{(4)})\right)\right] \\
S_{ij}&= \nabla^2 C_{ij} -2 R^{k\phantom{i}l}_{\phantom{k}i\phantom{l}j}C_{kl}+ 4\left((g_{(2)} g_{(4)}) -(g_{(4)} g_{(2)})\right)_{ij}+{1\over10} \big( \nabla_i \nabla _j B-g_{(0)ij} \nabla^2 B\big)\nonumber\\
&\phantom{=\ } +{2\over 5} g_{(2)ij}B + g_{(0)ij}\left(-{2\over 3} \tr g_{(2)}^3 -{4\over 15}(\tr g_{(2)})^3+{3\over 5} \tr g_{(2)} \tr g_{(2)}^2\right)
\end{align}
with the quantities $C_{ij}$ and $B$  defined by
\begin{align}
C_{ij} &=  g_{(4)ij} -{1\over 2} (g_{(2)}^2)_{ij} +{1\over 4}g_{(2) ij} \tr g_{(2)} +{1\over 8} g_{(0)ij} B\nonumber  \\
B&=\tr g_{(2)}^2-(\tr g_{(2)})^2
\end{align}
Note that the contraction of indices is performed with the inverse of $g_{(0)ij}$.
A general formula for the trace of the stress tensor follows from these definitions: 
\begin{align}
\langle T^{i}_{\phantom{i}i}\rangle&={3\, (2L)^5\over 8\pi G_N^{(7)}} \left( -{1\over 24} (\tr g_{(2)})^3 + {1\over 8} \tr g_{(2)} \tr g_{(2)}^2 -{1\over 6} \tr g_{(2)}^3 +{1\over 3} \tr g_{(2)}g_{(4)}\right)
\end{align}
Evaluating these formulae we find
\begin{align}
\langle T_{ij} \rangle\, dx^{i} dx^{j} &= \frac{2^4 L^5}{8\pi G_N^{(7)}}\, \frac{20 +9m_2^2-24 m_3}{160 } \left( ds^2_{AdS_3} - ds^2_{S^3} \right) 
\end{align}
Notice that the stress tensor is traceless, which reflects the fact there is no Weyl anomaly for $AdS_3 \times S^3$.
 After observing 
 \begin{align}
\frac{1}{8\pi G_N^{(7)}} &=  \frac{\textrm{Vol}(S^4_L)}{8\pi G_N^{(11)}}, \quad  \textrm{Vol}(S^4_L) = \frac{8\pi^2}{3}\, L^4
 \end{align}
 and using the definitions \eqref{eq:somedefinitions}, we  then  subtract off the contribution from the vacuum to obtain our final result:
 \begin{align}
\Delta \langle T_{i j} \rangle\, dx^{i} dx^{j} &= \frac{N^3}{160 \pi^3} \left(16 + 3m_2^2 - 8m_3\right)   \left( ds^2_{AdS_3} - ds^2_{S^3} \right) 
\end{align}

\section{Four-form field strength}
\setcounter{equation}{0}
\label{appast}

In this  appendix we present the formula for the four-form field strength
\begin{align}
F&= (f_1)^3 g_{1m} \, \omega_{AdS_3}\wedge e^m + (f_2)^3 g_{2m}  \, \omega_{S^3}\wedge e^m + (f_3)^3g_{3m} \, \omega_{\tilde S^3}\wedge e^m 
\end{align}
where $\omega_X$ denotes the volume form for a unit-radius space $X$. The $g_{I m}$ are related to derivatives of potentials $b_I$ via
\begin{align}
\label{bi}
(f_1)^3 g_{1w} = - \p_w b_1 / L^3
&= 2 (j_w^+ + j_w^-)
\no\\
(f_2)^3 g_{2w} = - \p_w b_2  / L^3&= - 2 (j_w^+ - j_w^-)
\no\\
(f_3)^3 g_{3w} = - \p_w b_3 / L^3 &= {1 \over 8} \, j_w^3
\end{align}
Since the  four-form field strength is related to the three-form potentials by $F_{(I)}= d C_{(I)}$, it follows from (\ref{bi}) that the potentials take the following form:
\begin{align}
C_{(1)}&= b_1 {1\over z^3}dz\wedge dt\wedge dl\nonumber\\
C_{(2)}&=b_2 \sin^2 \theta_1 \sin \theta_2\; d\theta_1\wedge \theta_2\wedge d\theta_3 \nonumber\\
C_{(3)}&=b_3 \sin^2 \psi_1 \sin \psi_2 \; d\psi_1\wedge d\psi_2\wedge d\psi_3
\end{align}

Next we review  the the expressions for the fields $j$  in (\ref{bi})   found in \cite{Bachas:2013vza}. 
The   currents can be expressed in a compact way by defining
\begin{align}
J_{w}&= {h\over L^3 (G+\bar G)} \left[ \bar G \left(G - 3 \bar G + 4 G \bar G^2\right) \p_w G + G \left(G + \bar G\right) \p_w \bar G \right]
\end{align}
and are given by
\begin{align}
j_w^+ &=
2  i   \,
J_{w}
\left((G - \bar G)^2 - 4 G^3 \bar G\right)    W^{-4}
\no\\
j_w^- &= 2  \, G J_{w}
\left(-2 G \bar G + 3 \bar G^2 - G^2 + 4 G^2 \bar G^2 \right)
 W^{-4}
 \no\\
j_w^3 &=
3  \p_w h \, {W^2 \over G (1 - G \bar G)}
-2 J_{w} \, {(1 + G^2) \over G  (1 - G \bar G)^2}
\end{align}
It is then straightforward to verify that   the potentials are  given by
\begin{align}
b_1&= {2 (G+\bar G) h \over 2 G \bar G + i (G-\bar G)}+ 2 \tilde h - 2 \Phi \nonumber \\
b_2&=-{2 (G+\bar G) h \over 2 G \bar G - i (G-\bar G)}+ 2 \tilde h + 2 \Phi  \nonumber \\
b_3&=- {(G+\bar G) h\over 4(G\bar G-1)} -\Phi 
\end{align}
Here, $\tilde h $ is the dual harmonic function to $h$ and satisfies
\begin{align}
i \partial_w h &=\partial_w \tilde h
\end{align}
With $h=-i L^3 (w-\bar w)$ as in (\ref{hgdefine}), one obtains
\begin{align}
\tilde h &= L^3 (w+\bar w)
\end{align}
Also, $\Phi$ is defined via 
\begin{align}
\label{phidefa}
\bar G \partial_w h & = \partial_w \Phi
\end{align}
Using $\partial_w h=-i L^3$ and $G$ given by (\ref{gdefa}) we solve  (\ref{phidefa}) to find
\begin{align}
\label{phidefb}
\Phi &= L^3 \sum_j (-1)^j \sqrt{ (w-\xi_j) (\bar w-\xi_j)}
\end{align}
Note that $\Phi$ is real,  hence the only thing that could be added is a constant,  corresponding to an ambiguity in the definition of the $b_I$.

\section{Calculation of the real line contribution to the on-shell action}
\setcounter{equation}{0}
\label{apprealx}

In this appendix we present details on the calculation of the contribution from the real line to the on-shell action. 
 To do this we have to expand the metric factors and $b_I$ in a power series in $y$ around $y=0$.  The important point is that the expansion of $G,\bar G$  differs in different intervals. Let us define
  \begin{align}
  \cI_0&= [ -\infty,\xi_1] \cup [\xi_2,\xi_3] \cup \cdots \cup   [\xi_{2n},+\infty] \nonumber\\
   \cI_+&= [ \xi_1,\xi_2] \cup [\xi_3,\xi_4] \cup \cdots \cup   [\xi_{2n-1},\xi_{2n}] 
    \end{align}
  For the Taylor series expansion of $G$ we have  
  \begin{align}
  \label{gexpan}
  G&= 
\left\{
\begin{array}{lc}
 0+g_1(x) y+ i g_2(x)  y^2+ g_3(x) y^3 +\ldots & \quad   x\in \cI_0  \\
i+  g_1(x) y+ i g_2(x)  y^2+ g_3(x) y^3+\ldots & \quad    x\in \cI_+
\end{array}
\right.
  \end{align}
  where 
  \begin{align}
  \label{gidefa}
  g_1(x)&=  \sum_j (-1)^j {1\over 2} {1\over |x-\xi_j|} \nonumber\\
  g_2(x)&= \sum_j (-1)^j {1\over 4} {{\rm sign}(x-\xi_j)\over |x-\xi_j|^2}\nonumber\\
  g_3(x)&= \sum_j (-1)^{j+1} {1\over 4} {1\over |x-\xi_j|^3}
  \end{align}
  For the calculation of $b_I$ we also need the Taylor series expansion of $\Phi$ defined in (\ref{phidefb}):
 \begin{align}
 \Phi &= L^3 \left( \phi_0(x) + \phi_2(x) y^2+ \phi_4(x) y^4+\ldots \right)
 \end{align} 
 where 
 \begin{align}
 \label{phi0defa}
 \phi_0(x)&= \sum_j (-1)^j |x-\xi_j| \nonumber\\
 \phi_2(x) &= \sum_j (-1)^j {1\over 2} {1\over |x-\xi_j|} \nonumber\\
 \phi_4(x)&=\sum_j  (-1)^{j+1} {1\over 8} {1\over |x-\xi_j|^3}
 \end{align}
 Note that $g_1=\phi_2$ and $g_3= 2 \phi_4$ which will be important in the expansion of the action.
 The combinations of metric functions appearing in (\ref{metrfaca}) can be expanded as follows:
 \begin{align}
  \label{metfac1}
 \left({f_2 f_3\over f_1}\right)^3 &=  \left\{
\begin{array}{lc}
{ L^3 (g_1^2-g_2)^2 \over g_1^2+g_2} y^3+ O(y^5) & x\in \cI_0 \\
&\\
-{ L^3 (g_1^2+g_2)^2 \over2( g_1^2+2g_2)} y^3+ O(y^5) & x\in \cI_+ \\
\end{array}
\right. 
 \end{align}
 and 
 \begin{align}
 \label{metfac2a}
 \left({f_1 f_3\over f_2}\right)^3 & =  \left\{
\begin{array}{lc}
-{ L^3 (g_1^2+g_2)^2 \over g_1^2-g_2} y^3+ O(y^5) & x\in \cI_0 \\
&\\
{4 L^3 \over g_1^4+ 3 g_1^2 g_2+ 2 g_2^2} {1\over y^3}+ O({1\over y}) & x\in \cI_+ \\
\end{array}
\right.
 \end{align}
 and
  \begin{align}
   \label{metfac3}
 \left({f_1 f_2\over f_3}\right)^3  &=  \left\{
\begin{array}{lc}
-{8 L^3 \over (g_1^4 -g_2^2) }{1\over y^3} +O({1\over y} ) & x\in \cI_0 \\
&\\
-{4 L^3 (g_1^2 + 2 g_2)^2 \over g_1^2 + g_2} y^3+ O(y^5) & x\in \cI_+ \\
\end{array}
\right.
 \end{align}
 The expansion of $b_I$ with $I=1,2,3$ works the same way, but there are some cancellations due to the relations of the expansion coefficients for $G$ and $\Phi$ mentioned above. We find
  \begin{align}
  b_1  &=  \left\{
\begin{array}{lc}
L^3 \left({4 g_1\over g_1^2- g_2} - 2 \phi_0 + 4 x  + O(y^2) \right) & x\in \cI_0  \\
&\\
L^3 \left({4 g_1\over g_1^2+ g_2} - 2 \phi_0 + 4 x  + O(y^2) \right) & x\in \cI_+  \\
\end{array}
\right.
 \end{align}
Since the subleading  term is of order $y^2$  as $y\to 0$ we find that $\partial_y(b_1^2)$ is of  order $y$. Also note that  the metric factor (\ref{metfac1}) is of order $y^3$ as $y\to 0$.  Thus we find that the contribution to the action coming from $b_1$ vanishes  at $y=0$ and hence does not contribute.

The Taylor expansion of $b_2$ is given by
 \begin{align}
b_2 & =  \left\{
\begin{array}{lc}
L^3 \left( -{4g_1\over g_1^2 +g_2} + 2 \phi_0 + 4 x  + O(y^2) \right) & x\in \cI_0  \\
 &\\
L^3 \left( (2\phi_0 + 4 x) + (g_1^3+ 3 g_1 g_2 -g_3) y^4 + O(y^6) \right) &
x\in \cI_+ 
\end{array}
\right.
 \end{align}
 It is important to note that for $x\in \cI_+$ we find that $\partial_y (b_2)^2$ will behave as $y^3$ as $y\to 0$ and together with the behavior of the metric factor (\ref{metfac2a}) produces a finite contribution to the action. 
 
 Similarly the Taylor expansion for $b_3$ is given by
 \begin{align}
b_3 &=  \left\{
\begin{array}{lc}
L^3 \left(-\phi_0 + \Big(g_1^3+ {g_3\over 2} \Big) y^4+ O(y^5) \right)& x\in \cI_0 \\
 &\\
L^3 \left( -{g_1\over g_1^2+ 2 g_2}-\phi_0  + O(y^2)\right) &
x\in \cI_+ 
\end{array}
\right.
 \end{align}
 In a similar manner as for $b_2$ we note that for $x\in \cI_0$ the $\partial_y (b_3)^2$ term will be of order $y^3$ which together with the behavior of the metric factor \eqref{metfac3} will produce a finite contribution to the action at $y=0$. 
 
 Summarizing we find that 
\begin{align}\label{giexp}
\lim_{y\to 0}{f_2^3 f_3^2\over 2 f_1^3} \partial_y (b_1^2) &=  0, \quad \quad x\in \mathbb{R} \nonumber \\
\lim_{y\to 0} {f_1^3 f_3^3\over 2 f_2^3} \partial_y (b_2^2)& =  \left\{
\begin{array}{lc}
0& x\in \cI_0 \\
 &\\
 {32 L^9 (g_1^3+ 3 g_1 g_2-g_3)(\phi_0+ 2x)\over (g_1^2+g_2)(g_1^2+ 2 g_2)} &
x\in \cI_+ 
\end{array}
\right. \nonumber \\
\lim_{y\to 0} {f_1^3 f_2^3\over 2 f_3^3} \partial_y (b_3^2) &=  \left\{
\begin{array}{lc}
{16 L^9(2 g_1^3+ g_3) \phi_0 \over (g_1^2-g_2)(g_1^2+g_2)}& \quad \quad\quad\;\;  x\in \cI_0  \\
 &\\
0&
\quad \quad\quad \;\;x\in \cI_+ 
\end{array}
\right.
 \end{align}

The integration region $\cI_0$ is cut off by the large $r_c$ cutoff and includes the intervals $[-r_c,\xi_1]$ and $[\xi_{2n},r_c]$ that are responsible for $r_c$ divergent terms. Using the large $|x|$ expansion one can show  using the Taylor series expansions of \eqref{giexp} for large arguments  that  the contribution from the integral is given from the large integration limits $x_{c,+}$ and $x_{c,-}$  by
\begin{align}\label{infiniteterm}
\int_{\xi_{2n}}^{x_{c,+}} dx  {f_1^3 f_2^3\over 2 f_3^3} \partial_y (b_3^2)|_{y=0} &=L^9 \left[ -16 x_{c,+}^3 + 12 m_2 x_{c,+}^2  + (16 -9 m_2^2 + 16m_3) x_{c,+} \right] + {\rm finite} \nonumber\\
\int_{x_{c,-}}^{\xi_1} dx  {f_1^3 f_2^3\over 2 f_3^3} \partial_y (b_3^2)|_{y=0} &=  L^9 \left[-16 |x_{c,-}|^3 - 12 m_2 |x_{c,-}|^2  + (16-9 m_2^2 + 16 m_3 ) |x_{c,-}| \right] +  {\rm finite} 
\end{align}
Using the fact that $x_{c,+}= r_c(0,\varepsilon)$ and $x_{c,-}= r_c(\pi, \varepsilon)$ together with the relation of the radial cut-off to the FG cut-off parameter $\varepsilon$ given in (\ref{rcutrel}),  one can extract the contributions of the $x$-integral  that are divergent with respect to the cut-off $\varepsilon$ as follows:
 \begin{align}
\label{apsxdiv}
I_{(x)}^{\textrm{div}}&= + {L^9\over 48 \pi G_N^{(11)}} \, \textrm{Vol}(S^3)^2\, \textrm{Vol}(AdS_3) \left(-\frac{256}{\varepsilon^6}+ \frac{80-12m_2^2+32m_3}{\varepsilon^2}  \right) 
\end{align}

\newpage


\providecommand{\href}[2]{#2}\begingroup\raggedright\endgroup

\end{document}